\documentclass[11pt]{article}

\usepackage[margin=0.9in]{geometry}
\usepackage[round,authoryear]{natbib}
\usepackage{amsmath,amsfonts,bm}

\def\eqref#1{equation~\ref{#1}}
\def\1{\bm{1}}

\DeclareMathAlphabet{\mathsfit}{\encodingdefault}{\sfdefault}{m}{sl}
\SetMathAlphabet{\mathsfit}{bold}{\encodingdefault}{\sfdefault}{bx}{n}

\usepackage{hyperref}
\usepackage{url}
\usepackage{booktabs}
\usepackage{amsmath,amssymb,amsthm}
\usepackage{graphicx}
\usepackage{xcolor}
\usepackage{microtype}
\usepackage{float}

\newcommand{\mosaic}{\textsc{Mosaic}}
\newcommand{\emptytype}{\mathtt{X}}
\newcommand{\masktype}{\mathtt{MASK}}
\newcommand{\indic}{\mathbb{I}}

\title{MOSAIC: Addressing Space for Local, Steerable, and Scalable Atomistic Generation}
\author{Ishan M. Nadkarni$^{1}$, Sebastien Hamel$^{2}$,\\
Flynn Walsh$^{2}$, and Fei Zhou$^{2}$\\[0.5em]
\small $^{1}$The University of Texas at Austin, Austin, TX, USA\\
\small $^{2}$Physical and Life Sciences Directorate,\\
\small Lawrence Livermore National Laboratory, Livermore, CA, USA\\[0.35em]
\small \texttt{ishan.nadkarni@utexas.edu}\\
\small \texttt{hamel2@llnl.gov, walsh40@llnl.gov, zhou6@llnl.gov}
}
\date{}
\hypersetup{hidelinks,hypertexnames=false,
  pdftitle={MOSAIC: Addressing Space for Local, Steerable, and Scalable Atomistic Generation},
  pdfauthor={Ishan M. Nadkarni, Sebastien Hamel, Flynn Walsh, Fei Zhou}}

\begin{document}
\raggedbottom
\maketitle

\begin{abstract}

Generative models are opening new possibilities in molecular discovery and
materials science. Realizing this potential, however, requires both local control over atomic placement and the ability to assemble large, heterogeneous structures.
We introduce \mosaic{}, a hybrid mesh--particle representation that partitions space into anchored regions containing atom slots. Each slot encodes occupancy, atomic species, and a continuous displacement from its anchor, providing persistent spatial addresses for local conditioning, hierarchical generation, and scalable assembly. Because \mosaic{} defines a representation rather than a model,
it can be paired with different generative backbones. With the molecular backbone held fixed, its local support and coordinate prior improve molecular connectivity and validity; the resulting model samples efficiently across molecular sizes and enables fine-grained spatial control. At larger material scales, the same representation enables generation of atomistic polycrystals conditioned on coarse spatial layouts. Most notably, locality turns large-scale generation into a bounded assembly problem, enabling structure generation with more than one million atoms.
Together, this establishes local spatial addressing as a practical foundation for steerable generation of matter across scales, from
individual molecules to large-scale structures of broader engineering significance.

\end{abstract}

\section{Introduction}
\label{sec:introduction}

Generative models now produce chemically plausible molecules
\citep{hoogeboom2022equivariant} and candidate periodic crystals
\citep{xie2022cdvae,jiao2023diffcsp,zeni2025mattergen}. For engineering
applications, however, a central requirement is to specify what should change,
where it should change, and what surrounding structure should be preserved.
Examples include replacing a chemical group while retaining a drug candidate's
molecular framework, redesigning a selected segment of a polymer or protein,
and introducing defects or modifying interfaces within a material
\citep{schneuing2024diffsbdd,wang2022scaffolding,lei2024grandcanonical}.
These tasks require local generation consistent with the surrounding atomic environment.

The same requirement extends beyond isolated molecules and small periodic
cells. Engineered boundaries can combine strength and ductility in metals
\citep{lu2009strengthening}, while grain-boundary chemistry governs cycling
stability in nickel-rich cathodes \citep{yan2018tailoring}. Modeling such systems
requires control over local atomic arrangements and their organization across
larger structures. This motivates a representation that lets users specify
spatial constraints and supports generation at increasing system sizes.
Unordered particle sets preserve continuous coordinates but do not inherently
provide persistent addresses for selected regions, including empty ones. Grids
provide spatial addresses but represent atoms indirectly, requiring particle
recovery after generation \citep{pinheiro2023voxmol}. We
therefore seek a representation that combines continuous particles with
explicit spatial addresses for local control and scalable assembly.

We introduce \mosaic{}, a hybrid mesh--particle representation built from
spatially anchored atom slots. Each slot stores occupancy, atomic species, and a
continuous displacement from its anchor. Occupied slots decode directly to
particles: the mesh organizes the generated state without quantizing final
coordinates. As in image generation, where layouts and control maps guide
content at specified locations \citep{li2023gligen,zhang2023controlnet},
persistent slot addresses allow conditions to act on selected regions.
Neighboring regions provide context so that local generation remains consistent
with its surroundings. The slots define the state
evolved by diffusion or flow matching, so \mosaic{} does not prescribe a
competing neural architecture.
Instead, it can build on mature equivariant and Transformer pipelines that
already model atomistic geometry effectively, while adding a spatial interface
for local control and assembly
\citep{satorras2021egnn,hoogeboom2022equivariant,joshi2025adit}.

We evaluate this representation in molecular generation and large-scale
polycrystal assembly. Our contributions are:
\begin{itemize}
    \item \textbf{Data Representation.} A hybrid mesh--particle representation
    combines continuous atomic coordinates with persistent spatial addresses,
    providing a common interface for local conditioning across existing
    generative backbones.
    \item \textbf{Empirical and Theoretical Support.} Matched-backbone
    ablations establish improvements in molecular connectivity and validity,
    while size-scaling and timing benchmarks demonstrate gains in
    scaling and sampling speed. Theoretical analysis bounds expected
    per-particle transport action independently of system size at fixed cell
    width and shows reduced permutation ambiguity under local support
    (Appendix~\ref{app:spatial_transport}).
    \item \textbf{Control and Assembly.} Spatial conditioning steers molecular
    functional groups and prescribed grain morphology and orientation;
    bounded autoregressive assembly extends local generation to nickel
    polycrystals containing more than one million atoms.
\end{itemize}

\begin{figure}[!t]
    \centering
    \includegraphics[width=\linewidth]{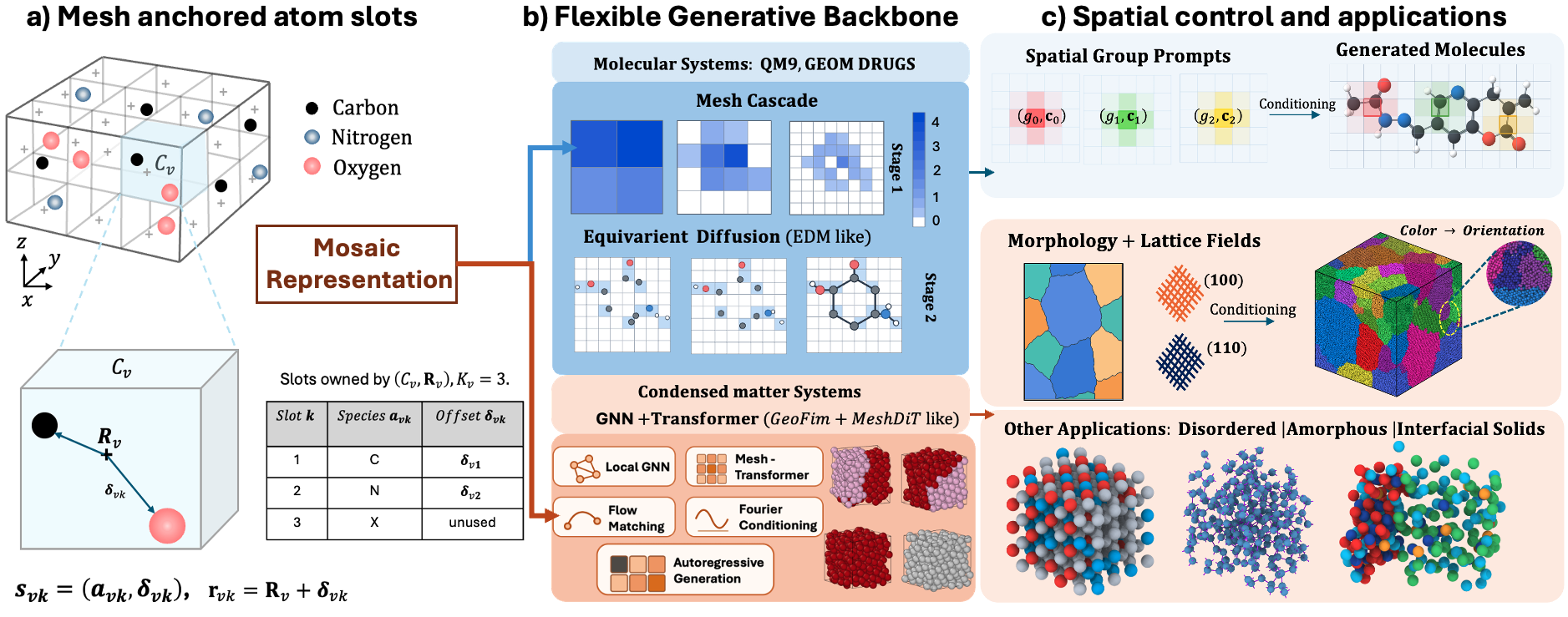}
    \caption{\mosaic{} representation and applications.
    (a) Each mesh region $C_v$ has an anchor $\mathbf{R}_v$ and
    $K$ candidate slots, each storing an atomic species or
    empty state $\emptytype$ and a continuous anchor-relative displacement.
    (b) The same representation supports distinct generative backbones for
    molecular and condensed-matter systems.
    (c) Spatial prompts steer molecular groups or grain morphology and lattice
    orientation, with applications spanning molecules, million-atom
    polycrystals, disordered alloys, amorphous materials, and interfaces.}
    \label{fig:overview}
\end{figure}

\section{Related work}
\label{sec:related}

\paragraph{Atomistic generative models.}
Equivariant diffusion established joint generation of molecular coordinates and
attributes \citep{hoogeboom2022equivariant}; latent, Bayesian-flow, and
Transformer variants followed
\citep{xu2023geoldm,song2024geobfn,reidenbach2025megalodon}.
Periodic-crystal diffusion models primarily address small cells
\citep{xie2022cdvae,jiao2023diffcsp,zeni2025mattergen}, while ADiT spans
molecules and periodic materials in a shared latent space \citep{joshi2025adit}.
FlowMol3 and Morph vary particle count through removable particles or graph
edits \citep{dunn2025flowmol3,franke2026morph}; molecular inpainting and
interaction profiles provide spatial control
\citep{schneuing2024diffsbdd,adams2024shepherd}. \mosaic{} makes locality part
of the particle representation, so spatial control is not tied to a particular
generative backbone.

\paragraph{Spatial and hierarchical representations.}
Voxel and neural-field methods make space explicit but encode atoms indirectly
through occupancy or density fields
\citep{pinheiro2023voxmol,kirchmeyer2025funcbind}.
Hierarchical generators obtain scalability from molecular fragments or coarse
chemical building blocks \citep{qiang2023hierdiff,fu2024mofdiff}, while MiAD
introduces removable fictitious atoms within particle diffusion
\citep{okhotin2026miad}. Uni-3DAR is closest to our spatial hierarchy,
tokenizing octree occupancy and fine-grained coordinates for autoregressive
generation \citep{lu2025uni3dar}. \mosaic{} instead makes locally anchored
particle slots the generated state: occupied slots decode directly to
continuous particles, and their addresses persist throughout diffusion or flow
generation, local conditioning, and boundary-aware assembly.

\paragraph{Large-scale atomistic generation.}
Size-conditioned models have generated amorphous silica structures containing
over 100,000 atoms \citep{yang2026amorphous}, while other approaches address
smaller disordered systems and individual grain boundaries
\citep{yong2024dismai,lei2024grandcanonical}. \mosaic{} targets prescribed
multigrain morphology, local crystallinity, and coherent assembly across many
generated blocks, producing nickel polycrystals containing over one million
atoms.

\section{MOSAIC}
\label{sec:method}

\subsection{Local mesh-anchored atom slots}
\label{sec:local_slots}

An atomistic configuration is an unordered marked point set
\begin{equation}
    \mathcal{X}=\{(\mathbf{r}_i,z_i)\}_{i=1}^{N},
    \qquad \mathbf{r}_i\in\Omega\subseteq\mathbb{R}^{3},
    \quad z_i\in\mathcal{E},
    \label{eq:particle_set}
\end{equation}
where $\mathcal{E}$ is the set of atomic species. We partition $\Omega$ into
regions $\{C_v\}_{v\in V}$ with anchors $\mathbf{R}_v$. This partition is a
spatial addressing scheme and need not coincide with the generated crystal
lattice. We use simple-cubic (SC), body-centered-cubic (BCC), or
face-centered-cubic (FCC) anchor grids, whose nearest-anchor regions form
regular partitions. Other anchor layouts can be used when a different
partition is better suited to the domain.

Region $v$ contains $K_v$ local slots. \mosaic{} can either predict this number
or keep it fixed. In the fixed-capacity setting, each anchor contains $K$
slots ($K_v=K$), so $M=|V|$ anchors provide $MK$ candidate
slots, including real and empty atom types; this follows local
grand-canonical and fictitious-particle formulations
\citep{lei2024grandcanonical,okhotin2026miad}. In the molecular model below,
Stage 1 instead predicts a local count $n_v$ and creates exactly $K_v=n_v$
real-atom slots at each finest-level cell.
All slots at an anchor share $\mathbf{R}_v$, while slot $k$ stores
\begin{equation}
    s_{vk}=(a_{vk},\boldsymbol{\delta}_{vk}=\boldsymbol{r}_i - \mathbf{R}_v),\qquad
    a_{vk}\in\mathcal{E}\cup\{\emptytype\},\qquad
    \boldsymbol{\delta}_{vk}\in\mathbb{R}^{3}.
    \label{eq:slot}
\end{equation}
Here $a_{vk}$ is either an atomic species or the physical empty state
$\emptytype$, and $\boldsymbol{\delta}_{vk}$ is an anchor-relative displacement of atom $i$, which is assigned to region $v=\arg \min_j |\boldsymbol{r}_i-\mathbf{R}_j|$, i.e.\ region $v$ is the Voronoi cell of $\mathbf{R}_v$.
Thus, one spatial address can represent multiple atoms, each with a distinct
species and displacement. In a fixed-capacity representation, unused slots take
the empty state $\emptytype$; in an exact-multiplicity representation, only the
predicted slots are instantiated without type $\emptytype$.
An occupied slot decodes directly to a continuous particle:
\begin{equation}
    D(\mathcal{S})=
    \left\{\left(\operatorname{wrap}_{\Omega}
    (\mathbf{R}_v+\boldsymbol{\delta}_{vk}),a_{vk}\right):
    a_{vk}\neq\emptytype\right\},
    \qquad
    N(\mathcal{S})=\sum_{v,k}\indic[a_{vk}\neq\emptytype].
    \label{eq:decode}
\end{equation}
The mesh therefore discretizes spatial ownership, not the final coordinate.
Figure~\ref{fig:overview}(a) illustrates this encoding: each anchor owns a small
set of slots, and occupied slots recover atom positions by adding a local
displacement to the anchor.
Atoms are assigned deterministically to nearby available slots during data
preparation, using minimum-image displacements for periodic systems. Full
assignment and boundary conventions are given in Appendix~\ref{app:method_details}.

This representation provides three useful biases. First, anchors replace a
global correspondence problem with smaller local ambiguities while preserving
the physical permutation symmetry of identical atoms. Second, local
displacements remain on an atomic length scale as the total structure grows.
Third, local capacity, whether represented by predicted multiplicities or by a
fixed bank with empty slots, permits variable atom count and allows conditions
to address regions that do not yet contain particles. The temporary diffusion
token $\masktype$ is distinct from the physical empty state $\emptytype$.

\subsection{Generative modeling with \mosaic{}}
\label{sec:mosaic_generation}

Let $\mathcal{O}=\{o_{vk}\}$ denote slot support, with
$o_{vk}=\indic[a_{vk}\neq\emptytype]$, and let $(A,\Delta)$ denote the species
and anchor-relative displacements of occupied slots. Conditioned on global or
spatial information $c$, the generative model factors as
\begin{equation}
    p(\mathcal{S}\mid c)
    =p_{\phi}(\mathcal{O}\mid c)\,
     p_{\theta}(A,\Delta\mid\mathcal{O},c),
    \label{eq:factorization}
\end{equation}
The factors may be modeled sequentially or jointly. For molecules, hierarchical
local counts first define $\mathcal{O}$, followed by equivariant flow generation
of $(A,\Delta)$. For polycrystals, a local graph--Transformer flow evolves
$(\mathcal{O},A,\Delta)$ jointly within each bounded block using spatial fields
and generated boundary context. Sections~\ref{sec:molecular_method}
and~\ref{sec:polycrystal_method} specify the two models.

\subsection{Molecular generation: hierarchy and steering}
\label{sec:molecular_method}

Let $\mathbf{n}^{(\ell)}=\{n_v^{(\ell)}:v\in V_\ell\}$ denote atom counts on
level $\ell$ of a spatial hierarchy, and let $\mathcal{O}(\mathbf{n}^{(L)})$
be the finest-level slot support. Given a requested atom count $N^\star$ and an
optional spatial prompt $\mathcal{P}$, Stage 1 refines a sparse hierarchy of
cell counts that defines the support for Stage 2. This parallels Uni-3DAR's
spatial hierarchy, but supplies a continuous equivariant molecular generator
\citep{lu2025uni3dar}:
\begin{equation}
    p(\mathcal{S}\mid N^\star,\mathcal{P})={}
    \underbrace{\prod_{\ell=0}^{L-1}
    p_{\phi}^{(\ell)}\!\left(
      \mathbf{n}^{(\ell+1)}\mid
      \mathbf{n}^{(\ell)},\mathcal{P}
    \right)}_{\text{Stage 1: hierarchical count and support}}
    \underbrace{p_{\theta}\!\left(
      A,\Delta\mid
      \mathcal{O}(\mathbf{n}^{(L)}),\mathcal{P}
    \right)}_{\text{Stage 2: atom types and coordinates}}.
    \label{eq:molecular_cascade}
\end{equation}
Setting $\mathcal{P}=\varnothing$ recovers unprompted, count-conditioned
generation.
Figure~\ref{fig:molecule_results}(a,b) visualizes the support-refinement and
atom-generation stages.
Stage 1 uses CountDiT, a spatial Diffusion Transformer whose weights are shared
across hierarchy levels. Finite-neighborhood attention uses relative cell-center
displacements and DiT/ADiT-style conditioning blocks
\citep{vaswani2017attention,peebles2023dit,joshi2025adit}. Each cell count is corrupted
independently by an absorbing masked diffusion process
\citep{austin2021d3pm},
\begin{equation}
    q_t(n_t\mid n_0)
    =\alpha(t)\,\delta_{n_0}(n_t)
     +\bigl[1-\alpha(t)\bigr]\,\delta_{\masktype}(n_t),
    \label{eq:count_corruption}
\end{equation}
and the network predicts clean count classes at masked cells. During sampling,
the model refines a root count across progressively finer meshes. At every
level, integer projection enforces
\begin{equation}
    n_{\mathrm{root}}^{(0)}=N^\star,
    \qquad
    \sum_{v\in\operatorname{ch}(u)}n_v^{(\ell+1)}=n_u^{(\ell)}
    \quad\Longrightarrow\quad
    \sum_{v\in V_L}n_v^{(L)}=N^\star.
    \label{eq:count_conservation}
\end{equation}
The finest-level count $n_v^{(L)}$ creates exactly that many slots in cell
$v$. Consequently, the Stage 2 support contains exactly $N^\star$ slots; the
retained zero-margin molecular model does not use the physical empty state
$\emptytype$ to add or remove atoms.
Stage 2 adapts EDM's EGNN backbone with DimeNet-style radial features, edge
attention, self-conditioning, anchor-relative coordinate updates, and a
dynamically rebuilt finite-cutoff graph
\citep{satorras2021egnn,hoogeboom2022equivariant,gasteiger2020dimenet,
chen2023analogbits}. It is trained on reference finest-level supports and
receives CountDiT-generated supports at inference. Writing
$\mathbf{x}_1=(A,\Delta)$ for the clean atom state and $\mathbf{x}_0$ for a
prior sample, training follows the conditional flow-matching path
\citep{lipman2023flowmatching}
\begin{equation}
    \mathbf{x}_t=(1-t)\mathbf{x}_0+t\mathbf{x}_1,
    \qquad t\in[0,1].
    \label{eq:molecular_flow_path}
\end{equation}
The prior uses Gaussian species channels and anchor-relative displacements
sampled uniformly within the corresponding finest-level cells.
Because Stage 2 predicts one atom for every slot created by Stage 1, it
preserves the requested atom count.

For GEOM-Drugs, molecules occupy a nonperiodic
cube of side length $25.6\,\text{\AA}$, with Stage 1 cell widths of
$12.8$, $6.4$, and $3.2\,\text{\AA}$. QM9 uses a separate finer hierarchy
reported in Appendix~\ref{app:molecular_method_details}. Both models generate
atomic species and continuous three-dimensional coordinates. Bonds, bond
orders, and formal charges are not generated and are reconstructed only for
evaluation.

The prompt $\mathcal{P}=\{(g_j,\mathbf{c}_j)\}_{j=1}^{J}$ specifies chemical
group types and desired centers. It induces one Gaussian channel per group,
evaluated on every Stage 1 mesh and again on the Stage 2 slots:
\begin{equation}
    h_g(\mathbf{x};\mathcal{P})=
    \sum_{j:g_j=g}
    \exp\!\left(-\frac{\|\mathbf{x}-\mathbf{c}_j\|_2^2}{2\sigma^2}\right),
    \qquad
    \widehat{\mathbf{y}}_s=
    \mathbf{y}_s(\varnothing)+\gamma_s
    \bigl[\mathbf{y}_s(\mathcal{P})-\mathbf{y}_s(\varnothing)\bigr],
    \quad s\in\{1,2\}.
    \label{eq:molecular_prompt_field}
\end{equation}
Here $\gamma_s$ independently controls classifier-free guidance
\citep{ho2022classifierfree} in the two stages. Dropping the prompt in Stage 1
retains the parent-count condition, so
both guidance branches satisfy Eq.~\ref{eq:count_conservation}. Further
architecture and training details are in
Appendix~\ref{app:molecular_method_details}.

Local support also simplifies the coordinate paths learned in Stage 2. To see
this, consider a slot in a three-dimensional cubic cell of width $h$, and let
$\boldsymbol{\delta}_i$ be its target offset and let
$\boldsymbol{\epsilon}_i\sim\operatorname{Unif}([-h/2,h/2]^3)$ be its source
offset. Along the coordinate path
$\boldsymbol{\delta}_{i,t}=(1-t)\boldsymbol{\epsilon}_i
+t\boldsymbol{\delta}_i$, the action satisfies
\begin{equation}
    \mathbb{E}[\mathcal{A}_i]
    =\mathbb{E}\!\int_0^1
      \|\dot{\boldsymbol{\delta}}_{i,t}\|_2^2\,dt
    =\mathbb{E}\|\boldsymbol{\delta}_i-\boldsymbol{\epsilon}_i\|_2^2
    \leq h^2.
    \label{eq:local_transport_action}
\end{equation}
The per-particle coordinate action is therefore bounded by $h^2$, independently
of the total molecule size. Spatial addresses also
restrict the $N!$ correspondences of an unstructured $N$-particle cloud to
$\prod_v n_v!$ within-cell permutations. If $n_v\leq K$, their logarithmic
ambiguity is at most $N\log K$, rather than $\log(N!)=\Theta(N\log N)$.
The mesh therefore acts as a blockwise transport preconditioner, replacing
globally ambiguous, system-scale motion with locally bounded motion. Prior
flow-matching analyses connect structured couplings to lower regression
variance and straighter generative paths
\citep{pooladian2023multisample,klein2023efm,song2023equifm}. The derivation,
global-center comparison, and assumptions are given in
Appendix~\ref{app:spatial_transport}.

\subsection{Polycrystal generation: morphology, lattice, and scale}
\label{sec:polycrystal_method}

Polycrystal generation separates mesoscale organization from local lattice
placement. We prescribe the morphology with
\begin{equation}
    M(\mathbf{x})=(m_1(\mathbf{x}),\ldots,m_G(\mathbf{x})),
    \qquad m_g(\mathbf{x})\geq0,
    \qquad \sum_{g=1}^{G}m_g(\mathbf{x})=1,
    \label{eq:grain_morphology}
\end{equation}
where $m_g(\mathbf{x})$ is the local fraction of prescribed grain state $g$.
The map specifies grain number, location, shape, and adjacency. Such maps may
be obtained from experimental
reconstruction, phase-field simulation, or synthetic microstructures
\citep{groeber2014dream3d,chen2002phasefield,quey2011polycrystals}. The Ni
grain-state bank and mesoscale statistics are derived from the VoroTop
high-temperature FCC nickel polycrystal
\citep{lazar2018vorotop,vorotop_examples}.

Morphology specifies where each grain state belongs but not how its lattice is
placed. For each state, reciprocal vectors $\mathbf{q}$ and complex
coefficients from a representative local environment define
\begin{equation}
    F_g(\mathbf{q})=
    \frac{\sum_j w_j\exp(-\mathrm{i}\mathbf{r}_j^\top\mathbf{q})}
         {\sum_j w_j},\qquad
    \boldsymbol{\psi}_{g,\mathbf{q}}(\mathbf{x})=
    \begin{bmatrix}\operatorname{Re}\\ \operatorname{Im}\end{bmatrix}
    \left[
    \frac{F_g(\mathbf{q})}{|F_g(\mathbf{q})|+\varepsilon}
    \exp(\mathrm{i}\mathbf{q}^{\top}\mathbf{x})
    \right].
    \label{eq:reciprocal_fields}
\end{equation}
The direction and magnitude of $\mathbf{q}$ encode lattice orientation and
spacing, while the normalized complex phase fixes atomic registry. The fields
are evaluated at fixed slot anchors and evolving particle positions; per-peak
features are embedded, amplitude pooled, and added to the slot representation.
Their phase is translated analytically between block origins. Complete peak
extraction and feature construction are given in
Appendix~\ref{app:reciprocal_details}.

\paragraph{Local flow architecture.}
GeoFiLM combines SchNet-style continuous-filter graph messages with
feature-wise linear conditioning \citep{schuett2017schnet,brockschmidt2020gnnfilm}.
Slot states are pooled into coarse spatial tokens, processed by a Transformer,
and broadcast back before joint prediction of species or empty occupancy and
anchor-relative displacements
\citep{vaswani2017attention,peebles2023dit,joshi2025adit}. Architecture and
training details are given in Appendix~\ref{app:ar_details}.

\paragraph{Bounded autoregressive assembly.}
\label{sec:polycrystal_scaling}

The condensed-matter generator operates at two nested scales: an inner
conditional flow generates one bounded block, while an outer autoregressive
traversal propagates generated boundary context across blocks.

The domain is partitioned into adjacent blocks generated in raster order by a
shared local model. Each block retains a fixed slot capacity, with every slot
predicting an atomic species or the physical empty state $\emptytype$; its
realized atom count is the number of nonempty slots. Let $B_t$ denote the
occupied slots in block $t$, $\mathcal{C}_t$ its morphology and reciprocal
condition, and $H_t$ its generated predecessor boundary context. Then
\begin{equation}
    p_\theta(B_{2:T}\mid B_1,\mathcal{C})
    =\prod_{t=2}^{T}p_\theta(B_t\mid H_t,\mathcal{C}_t).
    \label{eq:ar_blocks}
\end{equation}
System size therefore increases the number of AR steps while the spatial domain
processed per step remains bounded by one block and its predecessor context.

For Ni, each $16\,\text{\AA}$ block contains 1024 slots and conditions on
generated atoms within $6\,\text{\AA}$ of its previously completed negative
$x$, $y$, and $z$ faces. These boundary atoms remain fixed while the
nonoverlapping block interior is generated. One empirical block initializes the
sequence; subsequent blocks use the prescribed morphology, phase-aligned
reciprocal fields, a material-level density prior, and generated predecessor
boundary atoms. Full architecture, training, traversal, and sampling details
are given in Appendix~\ref{app:ar_details}. Appendices~\ref{app:local_structure},
\ref{app:polycrystal_metrics}, and~\ref{app:broader_condensed} report
morphology agreement, local crystal order, and tests on
disordered and amorphous structures.

\section{Results}
\label{sec:results}

The results span three capabilities enabled by the \mosaic{} representation:
molecular quality and scaling, local spatial control, and atomistic generation
far beyond single periodic crystals. Molecular benchmarks isolate the effect of
structured support and measure performance across sizes; nickel polycrystals test
whether local control and structural coherence survive assembly across extended
domains.

\subsection{Molecular generation}
\label{sec:molecules}

\paragraph{Evaluation protocol.}
We evaluate 10,000 QM9 samples \citep{ramakrishnan2014qm9} and compare \mosaic{}
with the released EDM and MegalodonFlow GEOM-Drugs models on the same 1,000
requested atom counts and random seeds
\citep{hoogeboom2022equivariant,axelrod2022geom,reidenbach2025megalodon}. Our
primary metric requires the complete distance-derived molecular graph to be
connected and chemically valid. Strict RDKit reconstruction and PoseBusters
checks provide complementary geometry diagnostics
\citep{buttenschoen2024posebusters}. Sampling protocols are in
Appendix~\ref{app:molecular_evaluation}; evaluator definitions, uncertainty
estimates, and model-native comparisons are in Appendix~\ref{app:molecular_quality}.

\paragraph{Conventional molecular quality.}
On QM9, \mosaic{} completes all 10,000 requests with $84.44\%$ strict
all-request validity. On matched GEOM-Drugs requests, it more than doubles
EDM's connected-valid rate but has lower connected-valid and conditional
PoseBusters pass rates than MegalodonFlow (Table~\ref{tab:molecular_quality}).
MegalodonFlow jointly generates coordinates, bonds, and formal charges, so it
provides released-system context rather than a controlled architectural
comparison. A controlled support-and-prior ablation is reported separately in
Table~\ref{tab:molecular_support_ablation_main}.

\begin{figure}[H]
    \vspace{-0.55em}
    \centering
    \includegraphics[width=0.92\linewidth,trim=7 8 9 10,clip]{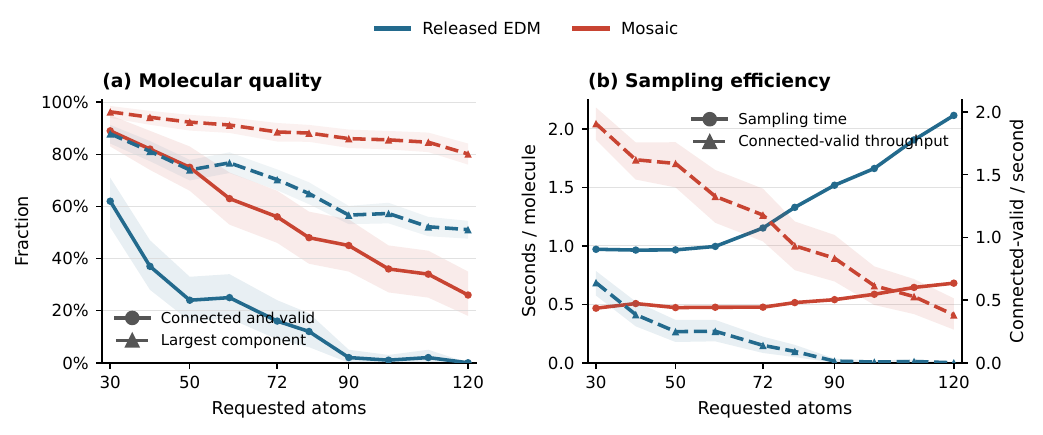}
    \vspace{-0.8em}
    \caption{GEOM-Drugs quality and sampling efficiency across requested atom
    counts under a shared full-graph evaluator (100 requests per model and
    count). All tested counts lie within \mosaic{}'s training range.
    Throughput is connected-valid molecules per sampling second.}
    \label{fig:geom_exact_size}
    \vspace{-0.45em}
\end{figure}

\paragraph{Accuracy and scaling.}
Across requests of 30--120 atoms, with 100 generations per model and size,
\mosaic{} retains higher connected-valid and largest-component fractions
throughout (Figure~\ref{fig:geom_exact_size}). At 120 atoms, EDM falls to zero
connected-valid molecules while \mosaic{} remains at $26\%$. This comparison
uses the same full-range \mosaic{} checkpoints as
Table~\ref{tab:molecular_quality} and tests size scaling within the training
range, not extrapolation beyond it. The trend is consistent with
Eq.~\ref{eq:local_transport_action}:
fixed-width support keeps coordinate transport local as molecule size grows.

\begin{table}[H]
    \centering
    \caption{Complete-model GEOM-Drugs \citep{axelrod2022geom} results on the
    same 1,000 requests under shared coordinate-only evaluators. Entries are
    percentages; PB-valid is conditional on strict RDKit validity. Full
    definitions, confidence intervals, and model-native metrics are in
    Appendix~\ref{app:molecular_quality}.}
    \label{tab:molecular_quality}
    \footnotesize
    \setlength{\tabcolsep}{3.5pt}
    \begin{tabular}{@{}lrrrrr@{}}
        \toprule
        Model & Atom stab. & Conn.-valid & RDKit valid & Connected & PB-valid \\
        \midrule
        EDM & 80.97 & 34.9 & 87.2 & 40.9 & 19.0 \\
        \mosaic{} & \textbf{85.33} & 77.2 & 97.4 & 80.0 & 54.1 \\
        MegalodonFlow & 82.47 & \textbf{84.7} & \textbf{99.4} & \textbf{96.0} & \textbf{89.8} \\
        \bottomrule
    \end{tabular}
\end{table}

\paragraph{Efficiency and memory.}
Using each model's native schedule on matched standard requests, \mosaic{}
provides $3.07\times$ the connected-valid throughput of MegalodonFlow, a
quality--speed tradeoff given the lower pass rates above. Across sampling
budgets, 100-step EDM achieves higher connected-valid throughput than the
fastest reported \mosaic{} setting, but at a much lower connected-valid rate. At comparable
latency, \mosaic{} achieves a higher connected-valid rate than EDM. Complete
timing, memory, and quality--speed results are in
Appendix~\ref{app:molecular_efficiency}.

\paragraph{Support-and-prior ablation.}
Here, ``local anchors'' denotes the molecular Stage-2 instantiation of
\mosaic{} used in this ablation; \mosaic{} is the underlying representation,
whereas the comparison changes how that representation supplies spatial
support and coordinate priors within the same backbone.
We train two Stage-2 particle generators from scratch while holding the
architecture, training protocol, and pretrained CountDiT Stage 1 fixed. The
global control collapses the predicted slots to one center and uses a Gaussian
coordinate prior; the local model retains the predicted cell anchors and uses a
uniform cell-wise prior. The local model raises connected-valid generation from
$24.3\%$ to $65.9\%$ with comparable sampling time
(Table~\ref{tab:molecular_support_ablation_main}). Complete metrics are in
Table~\ref{tab:geom_support_ablation}.

\begin{table}[H]
    \centering
    \caption{Controlled support-and-prior ablation on GEOM-Drugs. Complete
    metrics and confidence intervals are in
    Table~\ref{tab:geom_support_ablation}.}
    \label{tab:molecular_support_ablation_main}
    \footnotesize
    \setlength{\tabcolsep}{3pt}
    \begin{tabular}{@{}lrrrrr@{}}
        \toprule
        Setting & Atom stab. (\%) & Conn.-valid (\%) & RDKit valid (\%) & s/mol. & Conn.-valid/s \\
        \midrule
        Global center & 74.13 & 24.3 & 63.2 & \textbf{0.475} & 0.512 \\
        Local anchors & \textbf{83.44} & \textbf{65.9} & \textbf{97.3} & 0.507 & \textbf{1.299} \\
        \bottomrule
    \end{tabular}
\end{table}

\begin{figure}[!htb]
    \vspace{-0.35em}
    \centering
    \includegraphics[width=\linewidth,trim=4 0 4 2,clip]{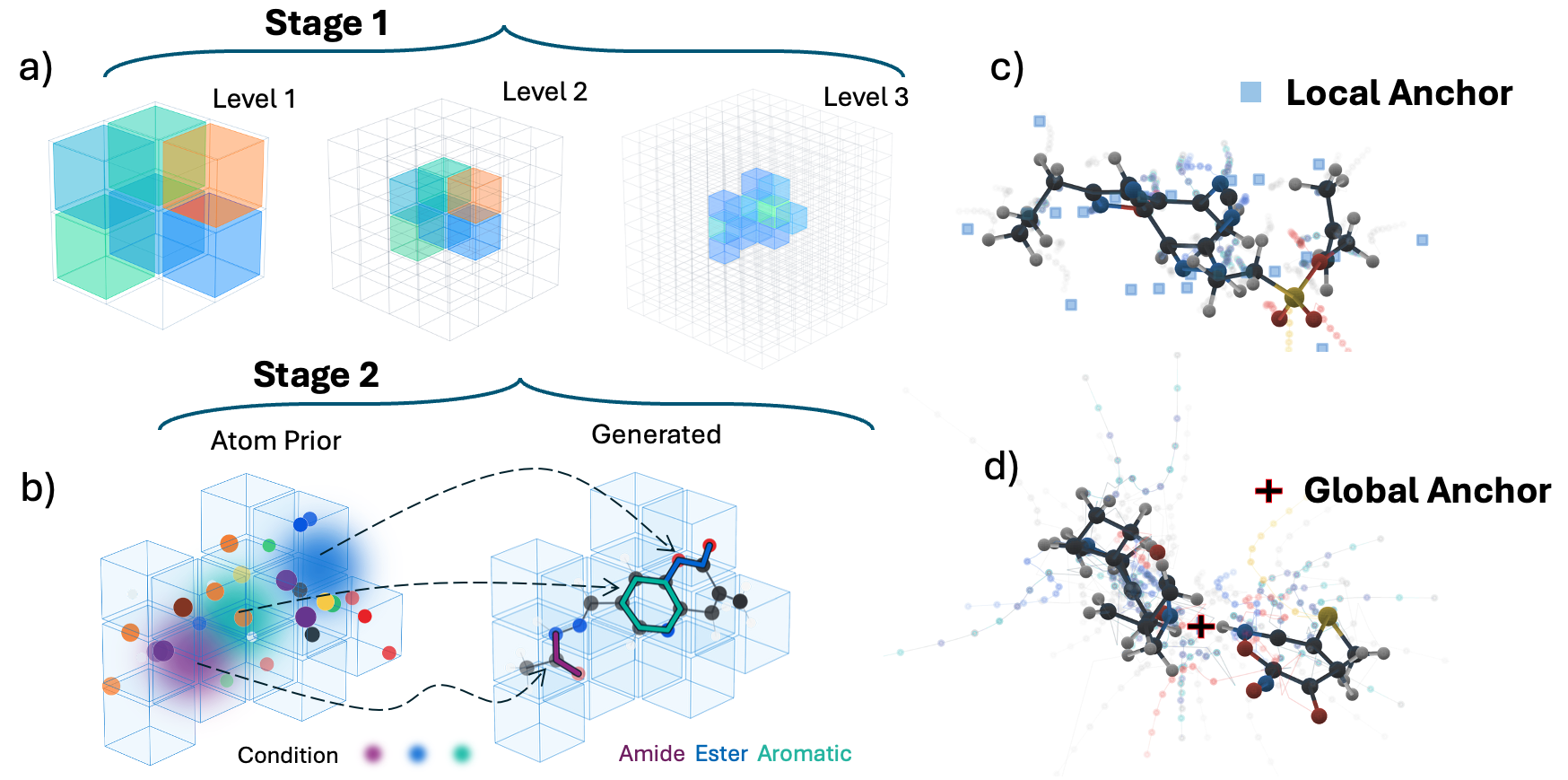}
    \caption{Molecular generation with structured support. (a) Count-conserving
    cascade. (b) Equivariant atom generation with steered groups highlighted. (c,d)
    Representative 55-atom outputs from the controlled comparison in
    Table~\ref{tab:molecular_support_ablation_main} show a connected local-support
    sample and a fragmented global-center sample.}
    \label{fig:molecule_results}
    \vspace{-0.45em}
\end{figure}

\paragraph{Local chemical and spatial steering.}
The same spatial fields steer both stages of the molecular cascade.
Figure~\ref{fig:molecule_results}(a,b) follows the requested amide, ester, and
aromatic groups from hierarchical support allocation through equivariant atom
generation. With Stage-2 guidance fixed at $\gamma_2=1$, increasing Stage-1
guidance from $\gamma_1=0$ to $2$ raises requested-group type $\mathrm{F}_1$
from $0.491$ to $0.667$ while preserving every requested atom count.
Appendix~\ref{app:molecular_steering} reports controlled Stage-2 and end-to-end
diagnostics; Figure~\ref{fig:molecular_sample_gallery} shows additional samples.

\subsection{Controlled polycrystal generation}
\label{sec:condensed}

\begin{figure}[t]
    \centering
    \includegraphics[width=\linewidth]{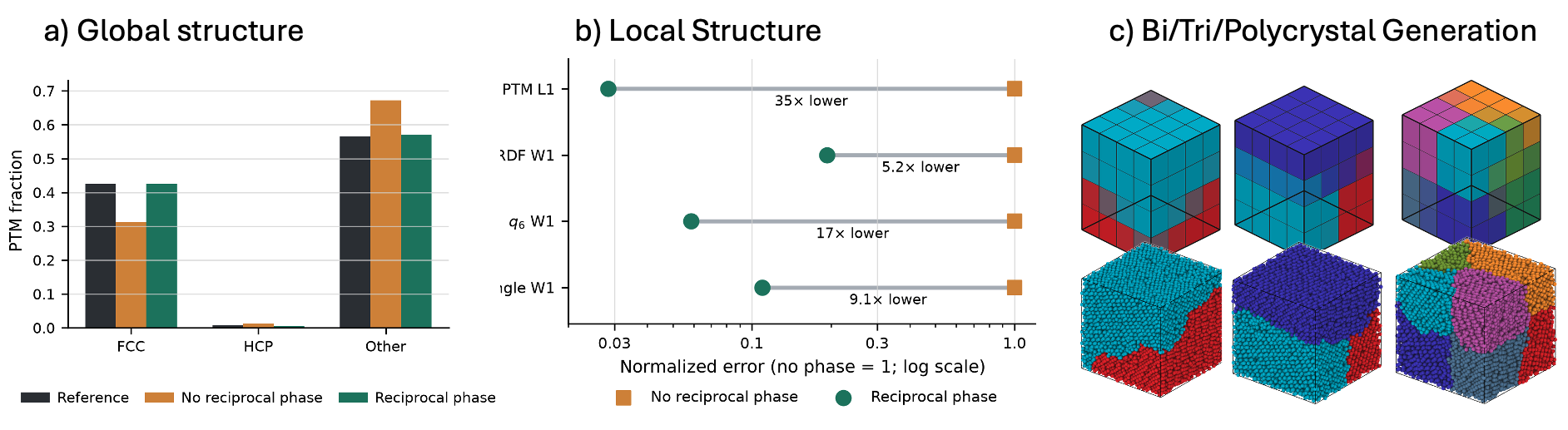}
    \vspace{-0.35em}
    \includegraphics[width=\linewidth]{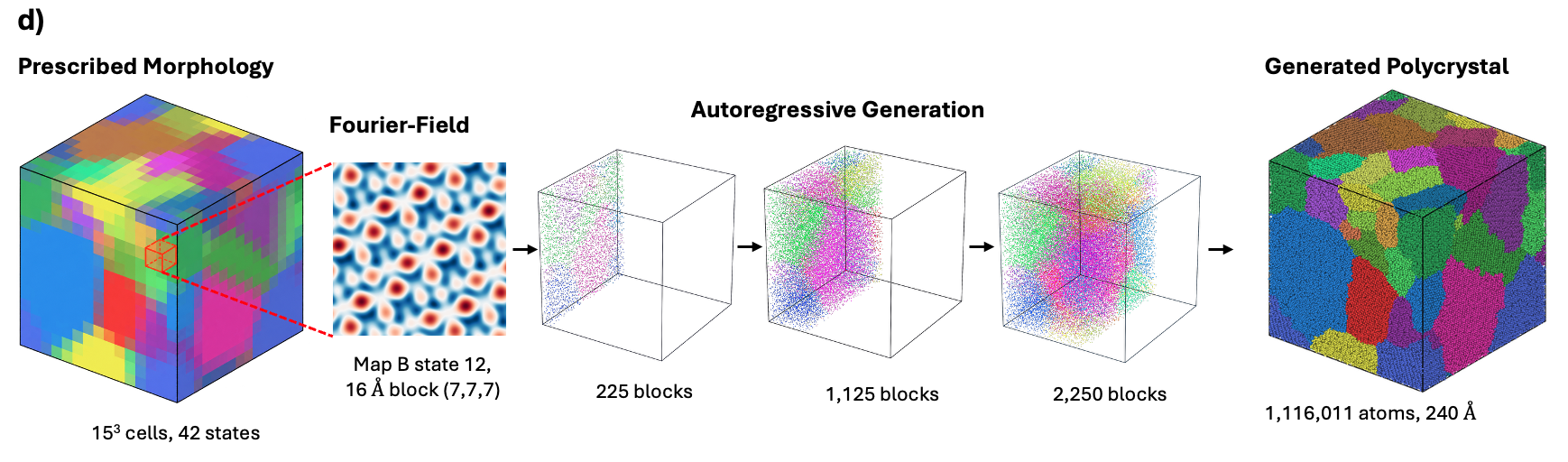}
    \caption{Reciprocal-field control and scalable polycrystal generation in
    nickel: (a) global structure, (b) local structural error, (c) controlled
    local morphologies, and (d) autoregressive million-atom assembly.}
    \label{fig:polycrystal_control}
\end{figure}

\paragraph{Ablation of reciprocal-phase conditioning.}
An important part of \mosaic{} is that information moves across scales in two
ways: the mesh carries spatial context, while reciprocal-space features carry
crystal orientation and phase. We test the second mechanism in a controlled
Ni environment. Adding reciprocal-phase features recovers the reference FCC
fraction and reduces PTM, RDF, $q_6$, and bond-angle errors by $5.2$--$35\times$
(Figure~\ref{fig:polycrystal_control}(a,b)). This controlled fit isolates the
representation's ability to encode lattice orientation and registry; the full
protocol is reported in Appendix~\ref{app:fourier_ablation}.

\paragraph{Local morphology control.}
\label{sec:local_morphology}

Coarse $4^3$ block maps define an oblique bicrystal, a tricrystal junction, and
an eight-grain Voronoi polycrystal.

The generated structures recover all requested states: 2/2, 3/3, and 8/8,
respectively (Figure~\ref{fig:polycrystal_control}(c)). This transfers editable
planar, junction, and multigrain morphologies into atomistic structure; complete
metrics are reported in Appendix~\ref{app:polycrystal_metrics}.

\paragraph{Million-atom nickel from coarse maps.}
We generate three prescribed 24-nm polycrystals from $15^3$ morphology maps
containing 41--44 grain states. Each begins from one atomistic seed block and
grows autoregressively: the model generates a bounded block from local
morphology and reciprocal fields while previously generated boundary atoms
provide context.

The completed structures in Figure~\ref{fig:ni_million_atom_ensemble} contain
1.09--1.12 million atoms;
Figure~\ref{fig:polycrystal_control}(d) shows an output conditioned on
42 prescribed orientation-state labels. Complete metrics are in
Appendix~\ref{app:polycrystal_metrics}.
Appendix~\ref{app:broader_condensed} further evaluates chemically disordered
alloys, amorphous silicon, and a five-species solid--solid interface, while
Appendix~\ref{app:local_structure} reports local-order diagnostics for Map B.

\section{Conclusions}
\label{sec:discussion}

\mosaic{} introduces a mesh--particle representation for atomistic generation
across a wide range of system sizes. The central idea is simple: atoms retain
continuous coordinates, but each atom is assigned to a local, spatially
anchored slot. This makes particle number and local conditions explicit and
keeps generation local as the system grows. In molecular benchmarks, this
representation improves connectivity across requested molecular sizes and
enables local chemical-group control. In condensed matter,
coarse grain maps and reciprocal-space fields guide the generation of extended
grain-boundary networks containing more than one million atoms.

The molecular and materials results use separately trained models, showing that
the representation is not tied to one architecture or generative objective.
This creates several opportunities. The slots can be combined with stronger
molecular and materials models, while mesh fields can encode charge,
composition, stress, density, or information from experiments and continuum
simulations. The same hierarchy may also provide a practical route to larger
biomolecules and proteins. More broadly, \mosaic{} extends atomistic generation
beyond isolated molecules and periodic cells
toward controllable, heterogeneous structures at experimentally and
technologically relevant length scales.

\bibliography{references}
\bibliographystyle{plainnat}

\clearpage
\appendix
\section{Additional method details}
\label{app:method_details}

\subsection{Slot assignment and boundary conventions}
The address set is
$\mathcal{I}=\{(v,k):v\in V,1\leq k\leq K_v\}$. The fixed-capacity
condensed-matter models use a uniform slot capacity $K_v=K$;
equivalently, each anchor is repeated $K$ times in the slot array. For molecular
generation, Stage 1 instead predicts $K_v=n_v$ and only those slots are
instantiated at the finest level. We construct regular anchor sets from SC, BCC,
or FCC grids at the desired spatial resolution. For periodic structures, the
grid is made commensurate with the simulation cell. These are anchor geometries
only and need not match the symmetry or phase of the generated material. The
same encoding applies to any user-defined anchor set equipped with local
ownership regions.

Each atom is assigned deterministically to an available slot near its position.
If atom $i$ is assigned to $(v,k)$, then
\begin{equation}
    a_{vk}=z_i,\qquad
    \boldsymbol{\delta}_{vk}=
    \operatorname{disp}_{\Omega}(\mathbf{r}_i,\mathbf{R}_v),
    \label{eq:app_encode}
\end{equation}
In the fixed-capacity parameterization, unused slots are set to
$(\emptytype,\mathbf{0})$. In the predicted-multiplicity molecular
parameterization, $K_v=n_v$ and there are no unused slots in the retained
zero-margin support. Assignment uses the cell containing the atom or the closest
available anchor while respecting local capacity. The displacement is an
ordinary relative vector for finite systems and a minimum-image vector for
periodic systems. Decoding uses the identity for finite systems and periodic
wrapping otherwise.

For a cubic cell of width $h$, anchor-relative coordinates normally lie within
$[-h/2,h/2]$ in each direction, independent of the total system size.
Distance-based capacity matching can relax this strict bound when a local cell
is full while retaining short displacements. Provided sufficient slot capacity,
encoding and decoding preserve the original species and coordinates without a
learned voxel-to-particle reconstruction step.

The physical empty state $\emptytype$ and diffusion state $\masktype$ have
different meanings. The former denotes a candidate slot that contains no atom.
The latter is a temporary absorbing categorical-noise token used during training
and is never decoded as a particle.

\subsection{Local transport and permutation reduction}
\label{app:spatial_transport}

We derive the coordinate-action bound in
Eq.~\ref{eq:local_transport_action} for the molecular Stage 2 model. Condition
on a finest-level support and consider slot $i$ assigned to cubic cell $v$ of
width $h$. Its clean and prior coordinates are
\begin{equation}
    \mathbf{x}_{1,i}=\mathbf{R}_v+\boldsymbol{\delta}_i,
    \qquad
    \mathbf{x}_{0,i}=\mathbf{R}_v+\boldsymbol{\epsilon}_i,
    \qquad
    \boldsymbol{\epsilon}_i\sim
    \operatorname{Unif}([-h/2,h/2]^3).
    \label{eq:app_local_coordinate_pair}
\end{equation}
For the linear conditional flow path, the target velocity is constant and the
kinetic action equals the squared endpoint displacement:
\begin{equation}
    \mathcal{A}_i
    =\int_0^1\|\dot{\mathbf{x}}_{t,i}\|_2^2\,dt
    =\|\mathbf{x}_{1,i}-\mathbf{x}_{0,i}\|_2^2
    =\|\boldsymbol{\delta}_i-\boldsymbol{\epsilon}_i\|_2^2.
    \label{eq:app_local_action}
\end{equation}
The source offset has zero mean and coordinate variance $h^2/12$. Therefore,
conditional on the target offset,
\begin{align}
    \mathbb{E}[\mathcal{A}_i\mid\boldsymbol{\delta}_i]
    &=\|\boldsymbol{\delta}_i\|_2^2
      +\mathbb{E}\|\boldsymbol{\epsilon}_i\|_2^2
      -2\boldsymbol{\delta}_i^\top
       \mathbb{E}[\boldsymbol{\epsilon}_i] \notag\\
    &=\|\boldsymbol{\delta}_i\|_2^2+\frac{h^2}{4}
      \leq h^2,
    \label{eq:app_local_action_bound}
\end{align}
where the last inequality uses
$\|\boldsymbol{\delta}_i\|_2^2\leq 3h^2/4$. Consequently,
$N^{-1}\mathbb{E}[\sum_i\mathcal{A}_i]\leq h^2$ for any $N$ as long as
source and target retain the same cell address. For the three-dimensional
GEOM-Drugs cells, $h=3.2\,\text{\AA}$ gives the conservative upper bound
$h^2=10.24\,\text{\AA}^2$ per particle. This is a worst-case geometric bound;
the observed mean can be substantially smaller.

For comparison, the matched global-center control represents every atom
relative to one anchor $\mathbf{R}_0$ and draws
$\mathbf{g}_i\sim\mathcal{N}(\mathbf{0},\sigma^2\mathbf{I}_3)$. Under its
independent source--target pairing,
\begin{equation}
    \frac{1}{N}\mathbb{E}[\mathcal{A}_{\mathrm{global}}]
    =3\sigma^2+
      \underbrace{\frac{1}{N}\sum_{i=1}^N
      \|\mathbf{r}_i-\mathbf{R}_0\|_2^2}_{M_2(\mathcal{X};\mathbf{R}_0)}.
    \label{eq:app_global_action}
\end{equation}
The local bound depends on cell width, whereas the second term in
Eq.~\ref{eq:app_global_action} depends on the spatial extent of the complete
molecule. For idealized compact structures growing at constant density in
three dimensions, this second moment scales as $O(N^{2/3})$, giving
$O(N^{5/3})$ total global action at fixed $\sigma$. With fixed bond lengths,
an ideal random-walk chain instead gives $O(N^2)$ total global action in
expectation, and an extended straight chain gives $O(N^3)$. In all three
cases, the local total action remains $O(N)$ at fixed cell width.

Spatial addressing also reduces correspondence ambiguity. An unstructured
cloud of $N$ exchangeable particles admits $N!$ assignments. Given local
occupancies $\{n_v\}$, an address-preserving coupling leaves only
\begin{equation}
    |\Pi_{\mathrm{local}}|=\prod_v n_v!,
    \qquad
    |\Pi_{\mathrm{global}}|=N!.
    \label{eq:app_permutation_counts}
\end{equation}
For $n_v\leq K$, the local bound and Stirling's approximation give
\begin{align}
    \log|\Pi_{\mathrm{local}}| &\leq N\log K,
    \label{eq:app_local_log_permutations} \\
    \log|\Pi_{\mathrm{global}}| &= N\log N-N+O(\log N).
    \label{eq:app_global_log_permutations}
\end{align}
Prior work connects
better couplings to lower regression variance and straighter, lower-cost flows
\citep{pooladian2023multisample,klein2023efm,song2023equifm}. The controlled
support-and-prior comparison in
Table~\ref{tab:molecular_support_ablation_main} shows the corresponding
improvement in molecular generation.

\subsection{Clean-state conditional flow objective}
\label{app:flow_objective}
Both particle renderers use the linear path
$\mathbf{x}_t=(1-t)\mathbf{x}_0+t\mathbf{x}_1$, where $\mathbf{x}_1$
concatenates clean species and anchor-relative displacement channels
\citep{lipman2023flowmatching}. The
network predicts the clean endpoint
$\widehat{\mathbf{x}}_1=f_\theta(\mathbf{x}_t,t,\mathbf{c})$ and minimizes
\begin{equation}
    \ell_i(t)=
    \left\lVert
    \frac{\mathbf{x}_{1,i}-\widehat{\mathbf{x}}_{1,i}}
         {1-\min(t,0.9)}
    \right\rVert_2^2,
    \qquad
    \mathcal{L}_{\mathrm{clean}}
    =\mathbb{E}\left[\frac{1}{|\mathcal{I}|}
    \sum_{i\in\mathcal{I}}\ell_i(t)\right].
    \label{eq:app_clean_state_loss}
\end{equation}
Species and coordinate channels are reweighted for each branch below. At
sampling time the corresponding velocity is
$\widehat{\mathbf{v}}_t=(\widehat{\mathbf{x}}_1-\mathbf{x}_t)/(1-t)$ and is
integrated with a fixed-step Euler solver.

\subsection{Molecular hierarchy and prompt fields}
\label{app:molecular_method_details}
\begin{figure}[t]
    \centering
    \includegraphics[width=\linewidth]{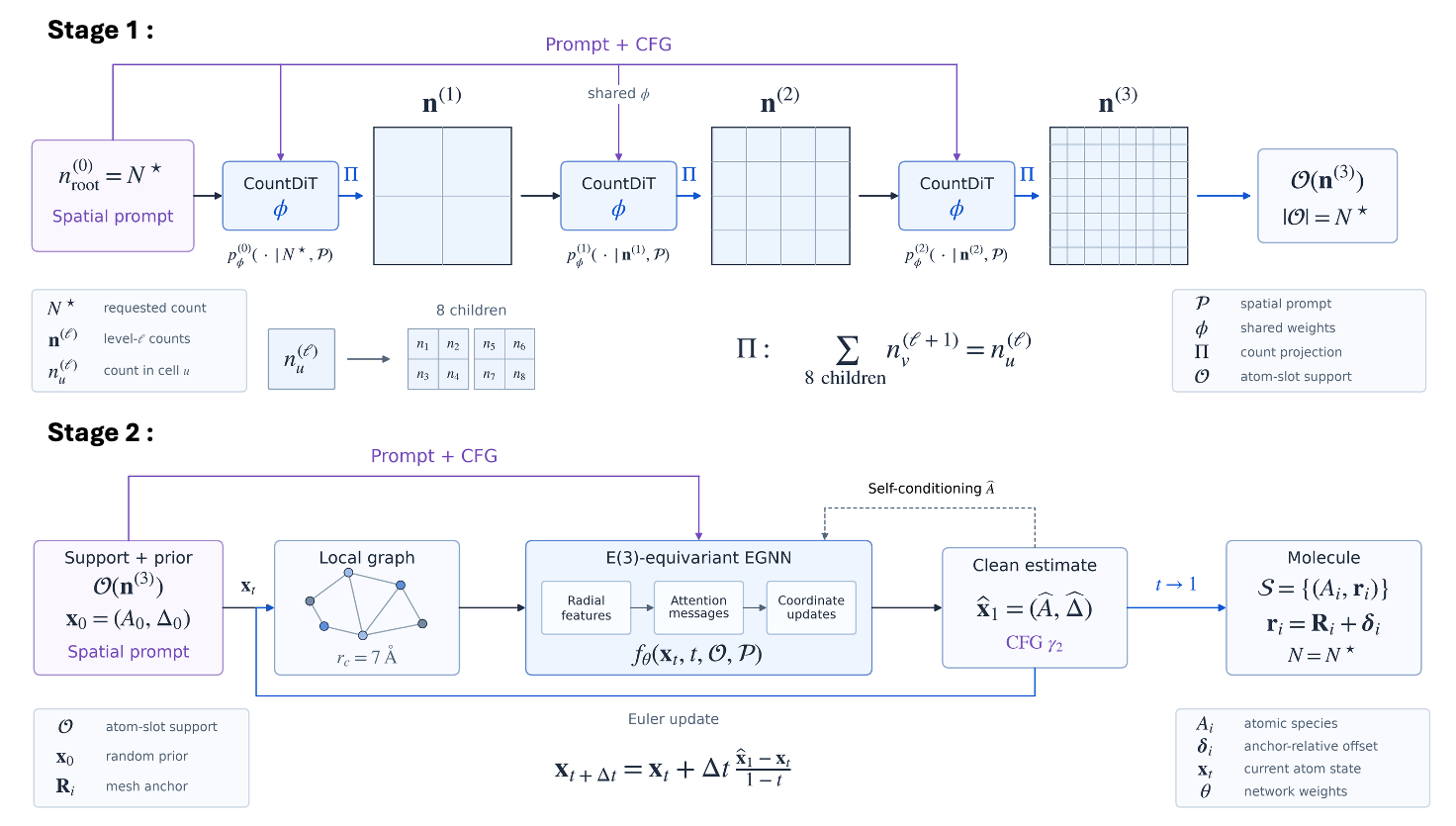}
    \caption{Molecular generation architecture. Stage 1 refines cell counts
    using a shared CountDiT, with integer projection $\Pi$ preserving each
    parent count and the requested total $N^\star$. Stage 2 generates species
    and continuous coordinates on the resulting support through equivariant
    flow matching. Spatial prompts guide both stages via classifier-free
    guidance. The grids show two-dimensional views of the three-dimensional
    GEOM-Drugs hierarchy.}
    \label{fig:app_molecular_training}
\end{figure}

Figure~\ref{fig:app_molecular_training} summarizes the molecular architecture
and spatial prompt conditioning.

\paragraph{Data and preprocessing.}
QM9 contains 130,831 structures with elements H, C, N, O, and F. GEOM-Drugs
uses corrected train, validation, and test splits of 243,458, 30,434, and
30,432 records before canvas filtering, with one deterministic energy-ranked
conformer per molecule and elements H, C, N, O, F, P, S, Cl, Br, and I. We
exclude 112 selected conformers that do not fit the fixed
$25.6\,\text{\AA}$ canvas. Random rotations are applied during Stage 2
training.

The QM9 model uses a $12.8\,\text{\AA}$ nonperiodic cube with Stage 1 cell
widths of $4.267$ and $2.133\,\text{\AA}$. GEOM-Drugs uses the three-level
hierarchy described in Section~\ref{sec:molecular_method}.

\paragraph{Stage 1: count diffusion.}
The Stage 1 CountDiT uses DiT/ADiT-style adaptive-layer-normalized Transformer
blocks \citep{peebles2023dit,joshi2025adit}. It has hidden width 256, eight
Transformer blocks, eight
attention heads, and an MLP expansion ratio of four. One set of weights is
shared across hierarchy levels. Level and diffusion-time embeddings identify
the current prediction task, while relative cell-center displacements enter
attention as a geometric bias within a finite spatial cutoff. Illegal count
classes are masked separately at each level. Let
$m_{bv}=\indic[n_{t,bv}=\masktype]$ and let $w_{bv}=0.1$ when the parent cell
is empty and $w_{bv}=1$ otherwise. Within a minibatch, define
$\omega_{bv}=m_{bv}w_{bv}/[1-\alpha(t_b)]$. The training objective is
\begin{equation}
    \mathcal{L}_{\mathrm{count}}
    =\sum_{\ell=0}^{L-1}
      \frac{\sum_b\sum_{v\in V_{\ell+1}}\omega_{bv}\,
      \mathrm{CE}\!\left(n_{bv},
      p_\phi^{(\ell)}(\cdot\mid\mathbf{n}_{t,b},
      \mathbf{n}^{(\ell)}_b)\right)}
      {\sum_b\sum_{v\in V_{\ell+1}}\omega_{bv}+\varepsilon}.
    \label{eq:app_count_objective}
\end{equation}
At sampling, predicted child counts are projected to nonnegative integers whose
sum equals the parent count. No additional conservation loss is used.

At the finest level, the instantiated molecular support is
\begin{equation}
    \mathcal{O}(\mathbf{n}^{(L)})
    =\{(v,k):v\in V_L,\ 1\leq k\leq n_v^{(L)}\},
    \qquad
    |\mathcal{O}(\mathbf{n}^{(L)})|
    =\sum_{v\in V_L}n_v^{(L)}=N^\star.
    \label{eq:app_molecular_support}
\end{equation}
Every instantiated slot becomes an atom. Thus the molecular pipeline
varies cardinality by predicting local slot multiplicities, not by classifying a
fixed bank of slots as occupied or empty.

\paragraph{Stage 2: molecular particle flow.}
Stage 2 adapts the EGNN backbone released with EDM
\citep{satorras2021egnn,hoogeboom2022equivariant} and trains it by conditional
flow matching \citep{lipman2023flowmatching}, rather than the EDM diffusion
objective itself. The
network has hidden width 256, nine message-passing layers, one invariant
sublayer per block, attention, and sum aggregation. Interatomic distances are
expanded in eight Bessel radial functions with a smooth polynomial envelope
following DimeNet \citep{gasteiger2020dimenet}, with a $7\,\text{\AA}$
interaction cutoff. Species priors are Gaussian with
scale $4$, while displacements are initialized uniformly inside their assigned
finest-level cells. Self-conditioning \citep{chen2023analogbits} is applied
with probability $0.5$. The
local neighbor graph is rebuilt from the noisy coordinates at each flow state,
and coordinate updates are bounded with a hyperbolic tangent. Stage 2 minimizes
Eq.~\ref{eq:app_clean_state_loss} over all instantiated slots and all species
and coordinate channels. It is trained on reference finest-level supports and
receives CountDiT-generated supports at inference. The retained configuration
uses no spare correction slots, so it cannot change the Stage 1 count or invoke
the physical empty state $\emptytype$.

The generated output is a coordinate-only atomistic structure: atomic species
and continuous three-dimensional coordinates are generated directly, whereas
bonds, bond orders, and formal charges are not model outputs. When required by
a molecular metric, these quantities are reconstructed from the generated
species and coordinates using the evaluation procedure in
Appendix~\ref{app:molecular_evaluators}.

The prompt field in Eq.~\ref{eq:molecular_prompt_field} uses
$\sigma=3.2\,\text{\AA}$ for all reported results. Repeated conditions of the
same type request multiple groups. Each cascade level has an independent
zero-initialized prompt projection because field amplitudes change with spatial
resolution. Prompt conditioning is dropped with probability $0.1$ during
training. Stage 1 retains the parent-count input when spatial conditioning is
absent, and
Stage 2 evaluates the field at atom-slot locations.

\paragraph{Optimization and training sequence.}
The base molecular models use AdamW \citep{loshchilov2019adamw} with zero
weight decay, gradient-norm
clipping at 1, a 2,000-update linear warmup followed by cosine decay to
$10^{-5}$, and exponential moving-average weights with decay $0.9999$.
GEOM-Drugs is initialized from the QM9 model for shared channels; tensors for
new elements and count classes are initialized separately, and optimizer state
is reset. The full-range GEOM models are warm-started from models trained up to
72 atoms and use a 75:25 mixture of natural count sampling and uniform sampling
from the 73--96, 97--128, and 129--181 atom ranges.
These full-range models are used in both
Table~\ref{tab:molecular_quality} and Figure~\ref{fig:geom_exact_size}. The
72-atom cutoff describes their initialization, not their final training range.

\begin{table}[H]
    \centering
    \caption{Molecular training budgets. Prompt adaptation uses a 500-update
    warmup; the final row lists trunk and prompt-projection learning rates.}
    \label{tab:molecular_training}
    \scriptsize
    \setlength{\tabcolsep}{4.2pt}
    \begin{tabular}{@{}llrrr@{}}
        \toprule
        Component & Data & Batch & Learning rate & Updates \\
        \midrule
        CountDiT & QM9 & 64 & $3\times10^{-4}$ & 200k \\
        Particle flow & QM9 & 64 & $1\times10^{-4}$ & 200k \\
        CountDiT & GEOM full range & 16 & $3\times10^{-4}$ & 80k \\
        Particle flow & GEOM full range & 8 & $1\times10^{-4}$ & 300k \\
        Prompt particle flow & GEOM $\leq72$ & 8 & $1\times10^{-4}$ & 120k \\
        Prompt CountDiT & GEOM $\leq72$ & 16 & $3\times10^{-5}/3\times10^{-4}$ & 13.2k \\
        \bottomrule
    \end{tabular}
\end{table}

Prompt training starts from the unprompted GEOM checkpoint pair. Stage 2 is
adapted first while Stage 1 is unchanged; Stage 1 is then adapted with Stage 2
fixed. The Stage 1 trunk and zero-initialized prompt projections use the two
learning rates in Table~\ref{tab:molecular_training}. This sequence preserves
the base generator while adding the conditional residual used by
classifier-free guidance \citep{ho2022classifierfree}.

\subsection{Reciprocal-field construction}
\label{app:reciprocal_details}
Let $\mathbf{r}_j\in[0,L]^3$ denote atoms from a representative crystalline
environment. To reduce boundary leakage, we apply the separable taper
\begin{equation}
    w_j=\prod_{d=1}^{3}\frac{1}{2}
    \left[1-\cos\!\left(\frac{2\pi r_{jd}}{L}\right)\right].
    \label{eq:app_reciprocal_taper}
\end{equation}
The weighted density is deposited on a $64^3$ grid, smoothed by a Gaussian with
standard deviation one grid cell, mean centered, and Fourier transformed. We
retain local intensity maxima with
$0.8\leq\lVert\mathbf{q}\rVert\leq10.0\,\text{\AA}^{-1}$, identify
$\mathbf{q}$ and $-\mathbf{q}$ as the same reciprocal axis, and apply
nonmaximum suppression. Each selected FFT proposal is refined by directly
maximizing the squared structure-factor magnitude in a bounded neighborhood.
The Ni renderer uses the eight strongest refined peaks; the matched local
ablation in Appendix~\ref{app:fourier_ablation} uses four.

For each retained vector $\mathbf{q}_m$, we compute
\begin{equation}
    F_m=
    \frac{\sum_jw_j\exp(-\mathrm{i}\mathbf{r}_j^\top\mathbf{q}_m)}
         {\sum_jw_j},
    \qquad
    A_m=|F_m|,
    \qquad
    u_m=\frac{F_m}{A_m+\varepsilon}.
    \label{eq:app_reciprocal_coefficient}
\end{equation}
Here $A_m$ measures crystalline coherence and $u_m$ fixes lattice registry.
At a query position $\mathbf{x}$, define the phase-aligned scalar and directional
fields
\begin{equation}
    z_m(\mathbf{x})=u_m e^{\mathrm{i}\mathbf{q}_m^\top\mathbf{x}},
    \qquad
    d_m(\mathbf{x})=\operatorname{Re}z_m(\mathbf{x}),
    \qquad
    \mathbf{g}_m(\mathbf{x})=
    \operatorname{Im}z_m(\mathbf{x})\widehat{\mathbf{q}}_m,
    \label{eq:app_reciprocal_fields}
\end{equation}
where $\widehat{\mathbf{q}}_m=\mathbf{q}_m/\lVert\mathbf{q}_m\rVert$.
For slot $i$, with anchor $\mathbf{R}_i$ and evolving position
$\mathbf{x}_{i,t}=\mathbf{R}_i+\boldsymbol{\delta}_{i,t}$, the complete peak
token is
\begin{equation}
\begin{split}
    \mathbf{t}_{im}=\bigl[&d_m(\mathbf{R}_i),\mathbf{g}_m(\mathbf{R}_i),
    d_m(\mathbf{x}_{i,t}),\mathbf{g}_m(\mathbf{x}_{i,t}),
    \operatorname{vec}_{\mathrm{sym}}(
    \widehat{\mathbf{q}}_m\widehat{\mathbf{q}}_m^\top),\\
    &\lVert\mathbf{q}_m\rVert/(2\pi), I_m,A_m\bigr]\in\mathbb{R}^{17},
    \label{eq:app_reciprocal_token}
\end{split}
\end{equation}
where $I_m$ is the FFT intensity normalized by the strongest retained peak and
$\operatorname{vec}_{\mathrm{sym}}$ stores the six unique components of the
axis dyadic. The dyadic is invariant to the arbitrary sign assigned to a
reciprocal axis. Peak order is removed by amplitude-weighted pooling,
\begin{equation}
    \boldsymbol{\phi}_i=
    \frac{\sum_{m=1}^{8}A_m\,\mathrm{MLP}(\mathbf{t}_{im})}
         {\sum_{m=1}^{8}A_m+\varepsilon},
    \qquad
    \mathbf{h}^{(0)}_i=\mathbf{h}^{\mathrm{real}}_i+
    \boldsymbol{\phi}_i.
    \label{eq:app_reciprocal_pooling}
\end{equation}
Thus the reciprocal condition enters the slot stream before GeoFiLM message
passing. For a block with global origin $\mathbf{o}$, evaluating the field in
local coordinates is equivalent to replacing $u_m$ by
$u_m\exp(\mathrm{i}\mathbf{q}_m^\top\mathbf{o})$. This analytic phase
translation preserves registry across block origins.

\subsection{Reciprocal-field ablations}
\label{app:fourier_ablation}

The direct ablation uses a 353-atom local environment selected from the nickel
polycrystal. Both models use the same mesh scaffold, GeoFiLM architecture,
5,000-update training budget, coordinate and occupancy noise scales, known
atom count, and sampling procedure. One model receives no reciprocal-phase
features; the second receives four reciprocal vectors and their associated
complex phase fields. This controlled single-environment experiment tests
whether the representation can express the lattice orientation and registry
when the relevant reciprocal information is available.

\begin{table}[H]
    \centering
    \caption{Matched reciprocal-phase ablation for a single nickel environment.
    All distances are evaluated against the same reference configuration;
    lower error is better.}
    \label{tab:fourier_matched}
    \small
    \setlength{\tabcolsep}{4.0pt}
    \begin{tabular}{@{}lrrrrr@{}}
        \toprule
        Condition & FCC & PTM L1 & RDF W1 (\AA{}) & $q_6$ W1 & Angle W1 (deg.) \\
        \midrule
        No reciprocal-phase features & 0.3127 & 0.2245 & 0.00998 & 0.0161 & 0.586 \\
        Reciprocal-phase features & 0.4246 & 0.00637 & 0.00193 & 0.00094 & 0.064 \\
        Reference & 0.4249 & -- & -- & -- & -- \\
        \bottomrule
    \end{tabular}
\end{table}

Table~\ref{tab:ni_conditioning_factors} reports a complementary
conditioning-factor analysis.
All conditions use the same trained autoregressive model, initialization
block, traversal order, diffusion seed, and 50-step sampler. The conditions
differ only in the descriptors supplied during generation. Daggered rows use
descriptors computed from the reference destination and therefore serve only
to identify which information is required by the model.

\begin{table}[H]
    \centering
    \caption{Nickel autoregressive conditioning analysis on a
    $1\times8\times8$ block sequence. The reference FCC fraction is 0.5623;
    lower PTM error is better. Daggers denote reference-derived diagnostic
    interventions.}
    \label{tab:ni_conditioning_factors}
    \scriptsize
    \setlength{\tabcolsep}{3.4pt}
    \begin{tabular}{@{}lrrrr@{}}
        \toprule
        Condition & FCC & PTM error & Count MAPE & Face-NN error \\
        \midrule
        Causal descriptors & 0.1336 & 0.4737 & 0.0772 & 0.1793 \\
        Reference reciprocal descriptor$^\dagger$ & 0.3175 & 0.2780 & 0.0324 & 0.1292 \\
        Reference global descriptor$^\dagger$ & 0.2248 & 0.3789 & 0.0670 & 0.1421 \\
        Reference atom count$^\dagger$ & 0.0902 & 0.5216 & 0.0000 & 0.2261 \\
        Reference reciprocal + global$^\dagger$ & 0.5107 & 0.0539 & 0.0055 & 0.0248 \\
        Reference peak geometry$^\dagger$ & 0.0664 & 0.5507 & 0.1395 & 0.1992 \\
        Reference coefficients$^\dagger$ & 0.0318 & 0.5885 & 0.1190 & 0.2356 \\
        All reference descriptors$^\dagger$ & 0.5074 & 0.0563 & 0.0000 & 0.0314 \\
        \bottomrule
    \end{tabular}
\end{table}

\subsection{Nickel source data}
\label{app:ni_source}
All nickel configurations used to construct the local blocks, grain-state
collection, and source-calibrated mesoscale priors originate from the VoroTop
high-temperature FCC nickel polycrystal example
\citep{lazar2018vorotop,vorotop_examples}. The distributed AtomEye file,
\texttt{unquenched.1300000.cfg}, contains $5{,}027{,}209$ Ni atoms in a
periodic $386.658\,\text{\AA}$ cubic cell. VoroTop describes the configuration
as an FCC nickel polycrystal at $1350\,\text{K}$, approximately $85\%$ of the
bulk melting temperature.

From this source we extracted the $12\times13\times13$ crop and
$16\,\text{\AA}$ blocks used for local modeling, then derived the empirical
grain-state collection and calibrated mesoscale statistics. The million-atom
rollouts use prescribed grain maps together with lattice and mesoscale
statistics calibrated from this source. They do not use reference atomic
coordinates for the generated domains; the material prior and single
initialization block are derived from the VoroTop configuration.

The source was segmented into 176 grains. Grain-count and size statistics were
estimated from 4096 periodic $240\,\text{\AA}$ windows, yielding a mean and
median of $42.33$ and $42$ grains, respectively. The three reported macrostates
use 41, 42, and 44 grain states, corresponding to central quantiles of this
distribution. Each prescribed $15^3$ map is a newly sampled periodic weighted
power diagram whose cell sizes and reciprocal states are drawn from the source
statistics rather than copied from a destination structure.

\subsection{Autoregressive model training and traversal}
\label{app:ar_details}
\begin{figure}[t]
    \centering
    \includegraphics[width=\linewidth]{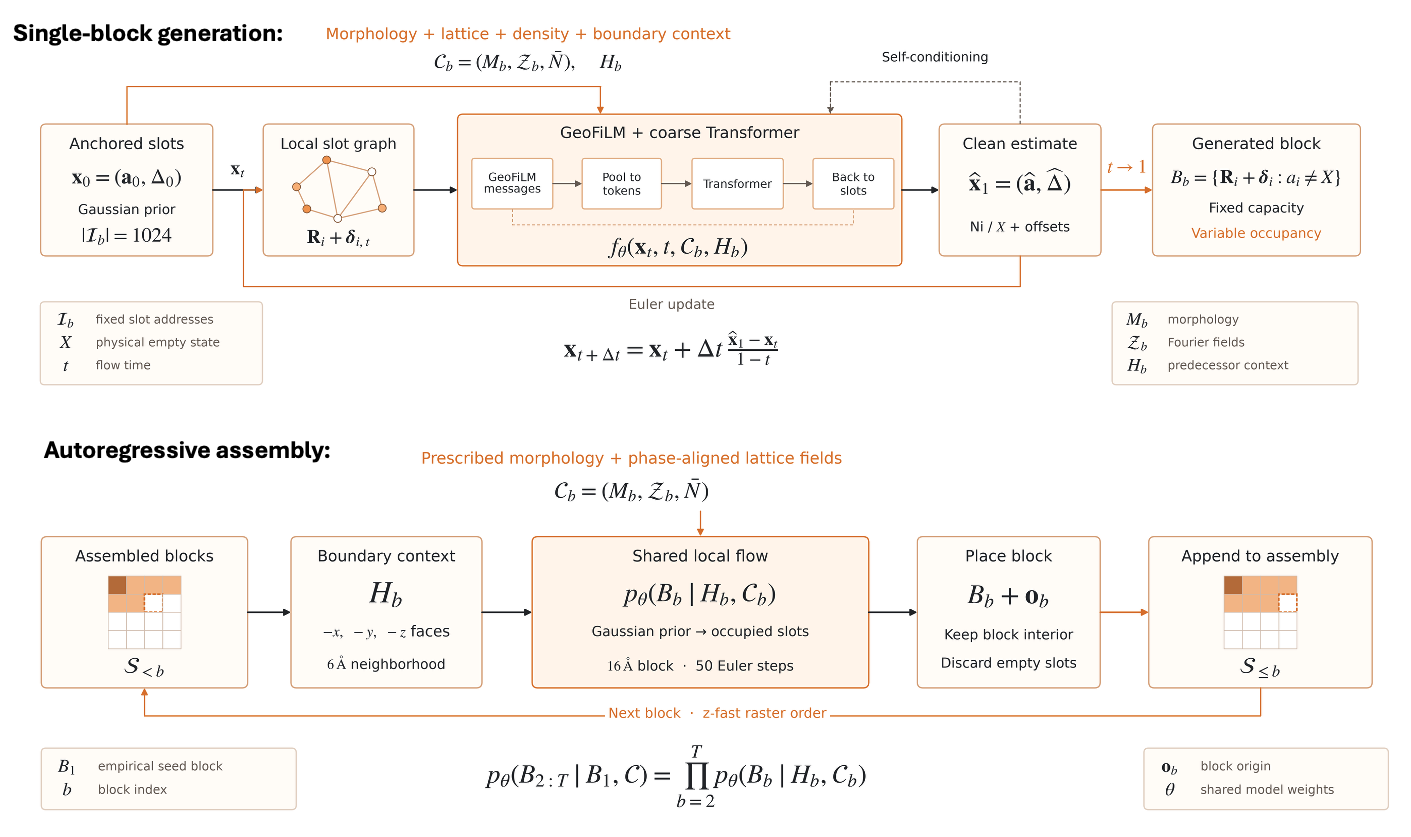}
    \caption{Nickel block generation and autoregressive assembly. Top: the
    local flow predicts species or the empty state $\emptytype$
    and continuous offsets on a fixed-capacity mesh. Bottom: the shared
    renderer advances in raster order using prescribed morphology,
    phase-aligned lattice fields, a density prior, and generated predecessor
    context. Here $b$ indexes blocks and $t$ denotes flow time.}
    \label{fig:app_polycrystal_training}
\end{figure}

Figure~\ref{fig:app_polycrystal_training} summarizes the local renderer and
autoregressive assembly.

The condensed-matter model uses a local atom-slot graph, geometric message
passing, a Transformer for coarse spatial tokens, and continuous flow matching
for coordinates and occupancy. Every block retains its full fixed
address set $\mathcal{I}_t$, and each slot predicts a type in
$\mathcal{E}\cup\{\emptytype\}$. Its realized atom count is therefore
\begin{equation}
    N_t=\sum_{(v,k)\in\mathcal{I}_t}
    \indic[a_{vk}\neq\emptytype].
    \label{eq:app_block_cardinality}
\end{equation}
This physical empty state is the condensed-matter cardinality mechanism and is
distinct from the predicted-multiplicity molecular support in
Eq.~\ref{eq:app_molecular_support}.

\paragraph{Architecture and data flow.}
Each slot receives noisy Ni or empty-species channels, a noisy anchor-relative
displacement, diffusion time, self-conditioning, reciprocal-field features,
and block and predecessor-face conditions. A dynamic slot graph applies
GeoFiLM messages, combining SchNet-style continuous-filter convolution with
GNN-FiLM modulation \citep{schuett2017schnet,brockschmidt2020gnnfilm}, using
radial, Cartesian direction, and moment features. Slot states are pooled into
coarse spatial tokens, processed by DiT/ADiT-style Transformer blocks
\citep{vaswani2017attention,peebles2023dit,joshi2025adit}, and broadcast back
to slots before clean species and displacement prediction. The
renderer uses width 256, two Transformer blocks with eight attention heads,
GeoFiLM message width 64, and one slot-feature layer. Boundary conditioning
comprises a six-dimensional face summary and 18-dimensional slot-local
geometric features. Parameters are shared across all blocks, and
self-conditioning \citep{chen2023analogbits} is applied with probability
$0.5$. The source prior is
independent Gaussian noise with scale $0.2$ for both the species channels and
the displacement channels, with displacement measured in \AA{}.

\paragraph{Loss.}
The species and displacement terms use the time normalization in
Eq.~\ref{eq:app_clean_state_loss}; displacement error is evaluated only on
occupied clean slots. Single-block pretraining uses only these two terms.
Boundary-conditioned adaptation uses the full objective
\begin{equation}
\begin{split}
    \mathcal{L}_{\mathrm{Ni}}={}&
    0.2\mathcal{L}_{\mathrm{type}}+\mathcal{L}_{\Delta}
    +0.01\mathcal{L}_{\mathrm{block\mbox{-}shell}}
    +0.02\mathcal{L}_{\mathrm{face\mbox{-}shell}}\\
    &+0.02\mathcal{L}_{\mathrm{seamNN}}
    +0.05\mathcal{L}_{\mathrm{contact}}.
    \label{eq:app_ni_objective}
\end{split}
\end{equation}
The block-shell and face-shell terms match smooth pair-distance histograms for
stitched blocks and cross-face pairs, respectively; the seam-NN term matches
sorted cross-face nearest-neighbor distances. For contact participation,
\begin{equation}
    s_i=\operatorname{sigmoid}\!\left(
    \frac{2.8\,\text{\AA}-d_i^{\mathrm{cross}}}{0.1\,\text{\AA}}
    \right),
    \qquad
    \mathcal{L}_{\mathrm{contact}}
    =\frac{1}{M}\left\lVert
    \operatorname{sort}(\mathbf{s}^{\mathrm{pred}})-
    \operatorname{sort}(\mathbf{s}^{\mathrm{ref}})
    \right\rVert_2^2.
    \label{eq:app_contact_loss}
\end{equation}

\paragraph{Training sequence.}
We use 128 raster-ordered training blocks from the
$12\times13\times13$ crop, with 16 validation and 16 test blocks. The first
stage is single-block lattice-conditioned pretraining from scratch, using the
known block count and target-derived reciprocal fields. Boundary-conditioned
adaptation warm-starts this model, retains those conditions, and adds clean
predecessor-face summaries and slot-local features. On half of training
examples, predecessor context is corrupted with $0.25\,\text{\AA}$ position
noise and 0.15 atom dropout. During autoregressive sampling, generated
predecessor atoms replace this teacher context.
Both stages use AdamW \citep{loshchilov2019adamw}, batch size four, zero weight
decay, and no learning-rate scheduler.

\begin{table}[H]
    \centering
    \caption{Nickel renderer training schedule.}
    \label{tab:ni_training}
    \scriptsize
    \setlength{\tabcolsep}{4.5pt}
    \begin{tabular}{@{}lrrl@{}}
        \toprule
        Stage & Learning rate & Selected update & Added condition \\
        \midrule
        Single-block pretraining & $1\times10^{-3}$ & 48,800 & Count and reciprocal field \\
        Boundary adaptation & $3\times10^{-4}$ & 57,440 & Predecessor faces \\
        \bottomrule
    \end{tabular}
\end{table}

\paragraph{Generation procedure.}
The following pseudocode gives the procedure used for the reported Ni
structures. Blocks follow raster order with the $z$ index changing fastest.

\begin{center}
\setlength{\fboxsep}{6pt}
\fbox{%
\begin{minipage}{0.94\linewidth}
\small
\textbf{Algorithm A.1: Causal autoregressive block assembly}\\[2pt]
\textbf{Input:} block lattice $\mathcal{B}$ with origins $\mathbf{o}_t$;
prescribed morphology $\mathcal{M}$ and reciprocal fields $\mathcal{Z}$;
renderer $f_\theta$; expected count $\bar N=356.3$; one empirical seed block
$\mathcal{S}_0$.\\
\textbf{Output:} assembled atom set $\mathcal{S}$.\\[2pt]
$\mathcal{S}\gets\operatorname{Place}(\mathcal{S}_0,\mathbf{o}_0)$\\
\textbf{for} $t\in\operatorname{Raster}(\mathcal{B}\setminus\{0\})$
\textbf{do}\\
\hspace*{1em}$\mathcal{C}_t\gets
\operatorname{Boundary}(\mathcal{S},\mathbf{o}_t;6\,\text{\AA})$
from available negative $x$, $y$, and $z$ faces\\
\hspace*{1em}$\mathbf{c}_t\gets
\operatorname{Encode}(\mathcal{M}(\mathbf{o}_t),
\mathcal{Z}(\mathbf{o}_t),\bar N,\mathcal{C}_t)$, including unavailable-face
indicators\\
\hspace*{1em}$(\mathbf{a}_t,\boldsymbol{\delta}_t)\gets
\operatorname{EulerFM}_{50}(f_\theta,\mathcal{I}_t,\mathbf{c}_t)$,
$|\mathcal{I}_t|=1024$, with self-conditioning\\
\hspace*{1em}$\widehat a_{vk}\gets
\arg\max_{e\in\mathcal{E}\cup\{\emptytype\}}a_{vke}$ for each slot
$(v,k)\in\mathcal{I}_t$\\
\hspace*{1em}$\mathcal{A}_t\gets
\{\mathbf{R}_{vk}+\boldsymbol{\delta}_{vk}:\,
\widehat a_{vk}\neq\emptytype\}$\\
\hspace*{1em}$\mathcal{S}\gets\mathcal{S}\cup
\operatorname{Interior}(\mathcal{A}_t+\mathbf{o}_t)$\\
\textbf{end for}\\
\textbf{return} $\mathcal{S}$
\end{minipage}}
\end{center}
Only one block and its predecessor boundary context are processed at each step,
so per-step memory does not grow with the assembled system size.

\section{Molecular protocols and baseline validation}
\label{app:molecular_evaluation}

\subsection{Models and sampling protocols}

The QM9 evaluation uses one retained \mosaic{} Stage-1 and Stage-2 checkpoint
pair to generate 10,000 samples, with 30 Stage-1 transitions per hierarchy
level and 100 Stage-2 transitions.

For GEOM-Drugs, all three systems receive the same 1,000 atom counts and
sampling seeds, with atom counts drawn from the retained training-split count
distribution. The only supplied condition is the total number of atoms;
neither composition nor reference geometry is provided.
Released EDM uses its native ancestral sampler with 1,000 reverse transitions
and 1,001 denoiser evaluations. \mosaic{} uses 32 reverse transitions on each of
three hierarchy levels and 50 Stage-2 transitions, totaling 146
step-equivalents. MegalodonFlow uses its released 100-step logarithmic
flow-matching sampler. Exponential moving-average weights are used in both
\mosaic{} stages. The exact-size study replaces the empirical count draw with 100
requests at each of 30, 40, 50, 60, 72, 80, 90, 100, 110, and 120 atoms.
Both Table~\ref{tab:molecular_quality} and Figure~\ref{fig:geom_exact_size}
use the same full-range GEOM-Drugs checkpoint pair: Stage 1 at 80,000 updates
and Stage 2 at 300,000 updates, trained as described in
Appendix~\ref{app:molecular_method_details}. Neither evaluation uses the earlier
models trained only on molecules with at most 72 atoms.

\subsection{Released EDM reproduction check}
Our released-checkpoint EDM results reproduce the behavior reported under
several published evaluator conventions (Table~\ref{tab:edm_reproduction}). In
particular, the $36.3\%$ published connected-valid value lies within our
$[31.9,37.8]\%$ interval. 
\begin{table}[H]
    \centering
    \caption{Comparison with published EDM results on GEOM-Drugs. Published values
    use the evaluator named in each row
    \citep{xu2023geoldm,le2024eqgat,reidenbach2025megalodon}.}
    \label{tab:edm_reproduction}
    \small
    \begin{tabular}{@{}lrrr@{}}
        \toprule
        Metric & Ours & Published & Difference \\
        \midrule
        Legacy atom stability & 80.97 & 81.3 & $-0.33$ pp \\
        Legacy validity & 95.5 & 92.6 & $+2.9$ pp \\
        Strict RDKit validity & 87.2 & 87.8 & $-0.6$ pp \\
        Strict full connectivity & 40.9 & 41.4 & $-0.5$ pp \\
        Full-graph connected-valid & 34.9 [31.9, 37.8] & 36.3 & within interval \\
        \bottomrule
    \end{tabular}
\end{table}

The apparent tension between legacy validity near $93\%$ and connected-valid
rates near $36\%$ is definitional. Legacy validity may sanitize the largest
fragment after disconnected components are removed; full-graph scoring requires
the entire generated structure to be one molecule. The original EDM study also
identifies disconnected components and long rings as characteristic GEOM-Drugs
failure modes \citep{hoogeboom2022equivariant}.

\section{Molecular efficiency and size scaling}
\label{app:molecular_efficiency}

\subsection{Matched native sampling}

Timing uses one MI300A, batch size four, one excluded warmup batch, and explicit
accelerator synchronization. It measures the sampling core only; checkpoint
loading, file output, bond reconstruction, PoseBusters, and all other chemistry
evaluation are excluded. Table~\ref{tab:geom_native_efficiency} reports the
native schedules.

\begin{table}[H]
    \centering
    \caption{Matched native-schedule sampling efficiency on the same 1,000
    standard GEOM requests. Connected-valid throughput uses the full-graph
    evaluator.}
    \label{tab:geom_native_efficiency}
    \small
    \begin{tabular}{@{}lrrrrr@{}}
        \toprule
        Model & s/molecule & samples/s & Conn.-valid/s & Peak MiB & Speedup \\
        \midrule
        Released EDM & 1.029 & 0.972 & 0.339 & 280.4 & $1.00\times$ \\
        \mosaic{} & 0.473 & 2.112 & 1.630 & 617.0 & $2.17\times$ \\
        MegalodonFlow & 1.597 & 0.626 & 0.530 & 1,098.8 & $0.64\times$ \\
        \bottomrule
    \end{tabular}
\end{table}

\subsection{Exact-size scaling}

\begin{figure}[H]
    \centering
    \includegraphics[width=0.92\linewidth]{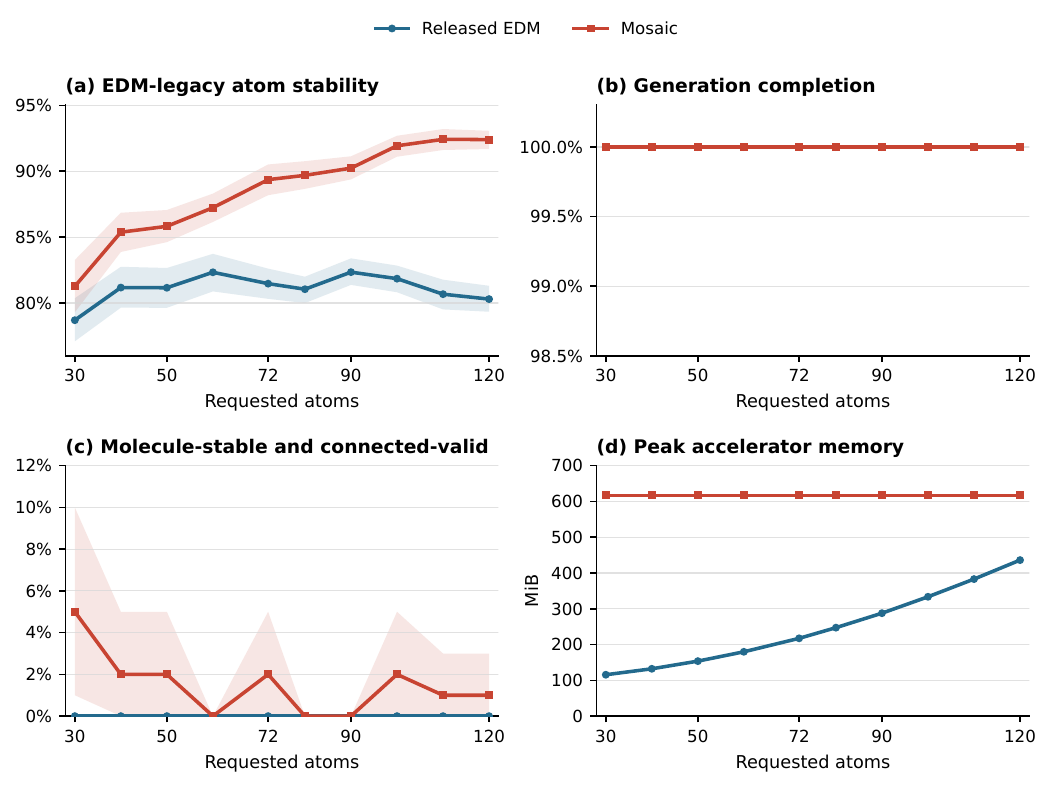}
    \caption{Additional exact-size diagnostics. The panels report legacy atom
    stability, completion, the joint stable and connected-valid rate, and peak
    accelerator memory. The current \mosaic{} implementation has a higher
    fixed memory floor, while EDM memory grows with requested atom count.}
    \label{fig:geom_exact_size_diagnostics}
\end{figure}

Table~\ref{tab:geom_exact_size_full} gives the complete exact-size ladder. The
connected-valid intervals are plotted in Figure~\ref{fig:geom_exact_size}; the
remaining stability, completion, and memory diagnostics appear in
Figure~\ref{fig:geom_exact_size_diagnostics}. The requested sizes lie within
the full-range GEOM training range; this study measures how quality and
computational cost vary with size, not extrapolation beyond the training range.

\begin{table}[H]
    \centering
    \caption{Complete GEOM exact-size scaling results. CV is full-graph
    connected-valid rate, LCF is mean largest-component atom fraction, and time
    is synchronized sampling seconds per molecule. Each entry uses 100
    requests.}
    \label{tab:geom_exact_size_full}
    \scriptsize
    \setlength{\tabcolsep}{3.8pt}
    \begin{tabular}{@{}rrrrrrrrr@{}}
        \toprule
        & \multicolumn{4}{c}{Released EDM} & \multicolumn{4}{c}{\mosaic{}} \\
        Atoms & CV & LCF & s/mol. & MiB & CV & LCF & s/mol. & MiB \\
        \midrule
        30  & 62 & 87.87 & 0.970 & 116 & 89 & 96.30 & 0.467 & 617 \\
        40  & 37 & 81.15 & 0.963 & 132 & 82 & 94.20 & 0.506 & 617 \\
        50  & 24 & 73.98 & 0.965 & 154 & 75 & 92.34 & 0.472 & 617 \\
        60  & 25 & 76.72 & 0.995 & 180 & 63 & 91.23 & 0.475 & 617 \\
        72  & 16 & 70.26 & 1.152 & 217 & 56 & 88.57 & 0.475 & 617 \\
        80  & 12 & 64.93 & 1.329 & 247 & 48 & 88.15 & 0.514 & 617 \\
        90  & 2 & 56.67 & 1.518 & 288 & 45 & 86.01 & 0.539 & 617 \\
        100 & 1 & 57.29 & 1.661 & 334 & 36 & 85.54 & 0.586 & 617 \\
        110 & 2 & 52.18 & 1.905 & 383 & 34 & 84.63 & 0.644 & 617 \\
        120 & 0 & 51.10 & 2.114 & 436 & 26 & 80.13 & 0.680 & 617 \\
        \bottomrule
    \end{tabular}
\end{table}

\subsection{Quality-speed frontier}

To separate native sampler choice from the quality-speed tradeoff, we evaluate
four schedules each for \mosaic{} and EDM on the same 1,000 standard requests.
At nearly matched latency, \mosaic{} with 10 hierarchy transitions and 25 Stage-2 transitions
reaches $72.7\%$ connected-valid at $0.284$ seconds per molecule, while EDM with
250 transitions reaches $37.5\%$ at $0.258$ seconds. The 100-step EDM setting
has the highest connected-valid throughput among the evaluated schedules,
but a lower connected-valid rate than every \mosaic{} setting
(Table~\ref{tab:geom_quality_speed}).

\begin{figure}[H]
    \centering
    \includegraphics[width=0.80\linewidth]{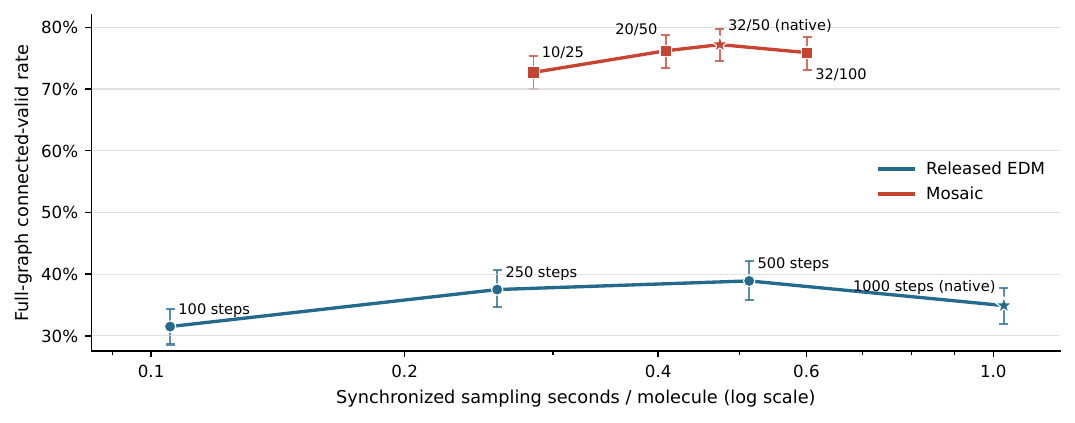}
    \caption{Quality-speed frontier on the same 1,000 GEOM requests. Bars show
    95\% bootstrap intervals and stars mark native schedules.}
    \label{fig:geom_quality_speed}
\end{figure}

\begin{table}[H]
    \centering
    \caption{Complete GEOM quality-speed schedule. Step-equivalents sum all
    hierarchy-level and Stage-2 transitions for \mosaic{}.}
    \label{tab:geom_quality_speed}
    \scriptsize
    \setlength{\tabcolsep}{3.5pt}
    \begin{tabular}{@{}llrrrrr@{}}
        \toprule
        Model & Budget & Step eq. & Conn.-valid & s/mol. & samples/s & Conn.-valid/s \\
        \midrule
        EDM & 100 & 100 & 31.5 [28.6, 34.4] & 0.105 & 9.49 & 2.99 \\
        EDM & 250 & 250 & 37.5 [34.6, 40.6] & 0.258 & 3.88 & 1.46 \\
        EDM & 500 & 500 & 38.9 [35.8, 42.1] & 0.513 & 1.95 & 0.76 \\
        EDM & 1,000 (native) & 1,000 & 34.9 [31.9, 37.8] & 1.029 & 0.97 & 0.34 \\
        \mosaic{} & 10/25 & 55 & 72.7 [70.0, 75.4] & 0.284 & 3.52 & 2.56 \\
        \mosaic{} & 20/50 & 110 & 76.2 [73.4, 78.7] & 0.408 & 2.45 & 1.87 \\
        \mosaic{} & 32/50 (native) & 146 & 77.2 [74.6, 79.8] & 0.473 & 2.11 & 1.63 \\
        \mosaic{} & 32/100 & 196 & 75.9 [73.1, 78.5] & 0.601 & 1.66 & 1.26 \\
        \bottomrule
    \end{tabular}
\end{table}

\section{Molecular quality evaluation}
\label{app:molecular_quality}

\subsection{Evaluator definitions and uncertainty}
\label{app:molecular_evaluators}

For the common coordinate-only comparison, all three models are evaluated from
generated atom types and coordinates without authoritative bond orders or formal
charges. MegalodonFlow's predicted graph and charges are evaluated separately.
We therefore keep the following evaluator families distinct.

\begin{itemize}
    \item \textbf{EDM legacy ($L$).} Bonds are inferred with the released EDM
    distance table. Legacy validity retains the largest connected fragment
    before sanitization.
    \item \textbf{Full graph ($F$).} The same distance-derived bonds are used,
    but every component is retained. Connected-valid requires the complete
    generated graph to be one connected, valid molecule.
    \item \textbf{Strict RDKit ($R$).} RDKit reconstructs bonds from coordinates
    using the first successful total charge in the fixed list
    $[0,-1,1,-2,2,-3,3]$. Explicit hydrogens and all connected components are
    retained.
    \item \textbf{Charge-marginalized corrected valency ($C$).} We apply the
    corrected valency definitions of \citet{nikitin2025geomrevisited} while
    searching the same fixed set of charge hypotheses. Because this common
    comparison does not use model-provided charge or bond labels, it is a
    diagnostic adapter, not an official corrected-GEOM score.
    \item \textbf{PoseBusters ($P$).} PoseBusters 0.3.1 applies its molecular
    configuration to strict-valid reconstructions
    \citep{buttenschoen2024posebusters}. PB-valid requires all seven reported
    intramolecular checks to pass.
\end{itemize}

All rates use every requested sample as the denominator unless explicitly
labeled \emph{among valid}. Confidence intervals use 2,000 bootstrap replicates
over requested samples.

The shared evaluation uses Python 3.13.2. Sampling uses PyTorch 2.7.0 with
ROCm 6.3 on one AMD Instinct MI300A. The full-graph evaluator uses ASE 3.29.0
and RDKit 2024.09.6; strict reconstruction uses RDKit 2026.03.6 and
PoseBusters 0.3.1.

\subsection{QM9 quality and geometric plausibility}

Table~\ref{tab:qm9_full_quality} gives the complete all-request QM9 summary.
Table~\ref{tab:qm9_posebusters} separates the PoseBusters pass fraction
among the 8,444 strict-valid molecules from the conservative fraction over all
requests.

\begin{table}[H]
    \centering
    \caption{Complete \mosaic{} quality summary on 10,000 QM9 requests. Rates
    and confidence intervals are percentages. Confidence intervals are not
    reported for uniqueness or novelty.}
    \label{tab:qm9_full_quality}
    \small
    \begin{tabular}{@{}lrr@{}}
        \toprule
        Metric & Value & 95\% interval \\
        \midrule
        Atom stability & 98.14 & [98.05, 98.23] \\
        Molecule stability & 80.68 & [79.87, 81.46] \\
        Strict RDKit validity & 84.44 & [83.70, 85.14] \\
        Full-molecule connectivity & 83.14 & [82.37, 83.88] \\
        Valid, connected, and stable & 77.23 & [76.36, 78.05] \\
        Uniqueness among valid & 99.22 & not bootstrapped \\
        Valid and unique over requests & 83.78 & not bootstrapped \\
        Novelty among unique valid & 83.17 & not bootstrapped \\
        \bottomrule
    \end{tabular}
\end{table}

\begin{table}[H]
    \centering
    \caption{PoseBusters checks for \mosaic{} on QM9. \emph{Among valid} uses
    successfully reconstructed strict-valid molecules; \emph{all requests}
    counts every invalid request as a failure.}
    \label{tab:qm9_posebusters}
    \small
    \begin{tabular}{@{}lrrr@{}}
        \toprule
        Check & Among valid & All requests & $N$ evaluated \\
        \midrule
        Connected atoms & 98.46 & 83.14 & 8,444 \\
        Bond lengths & 99.94 & 84.39 & 8,444 \\
        Bond angles & 99.93 & 84.38 & 8,444 \\
        Ring planarity & 100.00 & 84.44 & 8,444 \\
        Double-bond planarity & 97.41 & 82.25 & 8,444 \\
        Internal energy & 92.81 & 78.36 & 8,443 \\
        Steric clashes & 99.54 & 84.05 & 8,444 \\
        \bottomrule
    \end{tabular}
\end{table}

For broader context, Table~\ref{tab:qm9_published_context} collects published
QM9 values. Protocol labels are essential: the published EDM and GeoLDM rows
use the historical EDM evaluator, ADiT uses its native protocol, and \mosaic{}
uses strict all-request RDKit reconstruction.

\begin{table}[H]
    \centering
    \caption{Published QM9 results under different evaluation protocols. All
    values are percentages and published baselines are author-reported
    \citep{hoogeboom2022equivariant,xu2023geoldm,joshi2025adit}.}
    \label{tab:qm9_published_context}
    \small
    \setlength{\tabcolsep}{4pt}
    \begin{tabular}{@{}llrrrrr@{}}
        \toprule
        Model & Protocol & Atom stab. & Mol. stab. & Valid & Unique & Valid $\times$ unique \\
        \midrule
        EDM & legacy & 98.70 & 82.00 & 91.90 & n/r & 90.70 \\
        GeoLDM & legacy & 98.90 & 89.40 & 93.80 & n/r & 92.70 \\
        ADiT (QM9-only) & native & n/r & n/r & 96.02 & 97.76 & n/r \\
        \mosaic{} & strict RDKit & 98.14 & 80.68 & 84.44 & 99.22 & 83.78 \\
        \bottomrule
    \end{tabular}
\end{table}

\subsection{Matched GEOM-Drugs evaluation}

For model-native context, the original MegalodonFlow evaluation reports
$99.9\%$ atom stability, $98.8\%$ molecule stability, and $94.4\%$
valid-and-connected generation over 5,000 samples with its 100-step flow
sampler \citep{reidenbach2025megalodon}. A subsequent corrected reevaluation
reports $98.7\pm0.4\%$ molecule stability and $94.8\pm0.3\%$
valid-and-connected generation \citep{nikitin2025geomrevisited}. These
model-native values include predicted bonds and formal charges and are therefore
reported separately from the common coordinate-only evaluation below.

Table~\ref{tab:geom_full_quality} reports every primary evaluator family on the
matched standard-distribution requests. The charge-marginalized rows are kept
separate from the official corrected-GEOM benchmark because the generated
coordinates do not carry authoritative formal charges.

\begin{table}[H]
    \centering
    \caption{Complete matched GEOM-Drugs quality results. Brackets are
    requested-sample bootstrap 95\% intervals. PB-valid is among strict-valid
    molecules; strict uniqueness is among strict-valid molecules.}
    \label{tab:geom_full_quality}
    \small
    \setlength{\tabcolsep}{3.5pt}
    \resizebox{\linewidth}{!}{%
    \begin{tabular}{@{}lrrr@{}}
        \toprule
        Metric & Released EDM & \mosaic{} & MegalodonFlow \\
        \midrule
        Atom stability$^{L}$ & 80.97 [80.48, 81.43] & 85.33 [84.91, 85.78] & 82.47 [81.94, 82.97] \\
        Validity with fragment retention$^{L}$ & 95.5 [94.2, 96.7] & 98.2 [97.4, 99.0] & 97.5 [96.5, 98.4] \\
        Full-graph connectivity$^{F}$ & 36.9 [33.9, 40.0] & 78.4 [75.8, 80.9] & 86.6 [84.4, 88.7] \\
        Full-graph connected-valid$^{F}$ & 34.9 [31.9, 37.8] & 77.2 [74.6, 79.8] & 84.7 [82.4, 86.9] \\
        Strict RDKit validity$^{R}$ & 87.2 [85.1, 89.3] & 97.4 [96.4, 98.4] & 99.4 [98.9, 99.8] \\
        Strict connectivity$^{R}$ & 40.9 [37.8, 44.0] & 80.0 [77.4, 82.6] & 96.0 [94.8, 97.2] \\
        Charge-marginalized stability$^{C}$ & 83.9 [81.6, 86.2] & 96.8 [95.6, 97.9] & 99.4 [98.9, 99.8] \\
        Charge-marginalized valid and connected$^{C}$ & 40.9 [37.8, 44.0] & 80.0 [77.4, 82.6] & 96.0 [94.8, 97.2] \\
        PB-valid$^{P}$ & 19.00 & 54.11 & 89.83 \\
        Strict uniqueness & 100.0 & 100.0 & 100.0 \\
        \bottomrule
    \end{tabular}
    }
\end{table}

The individual PoseBusters outcomes in Table~\ref{tab:geom_posebusters} show
that the difference is not confined to graph connectivity. \mosaic{} also
improves bond geometry, double-bond planarity, internal energy, and steric-clash
rates under the same reconstruction procedure.

\begin{table}[H]
    \centering
    \caption{Detailed PoseBusters results on matched GEOM-Drugs requests.
    Each model column gives the pass rate among strict-valid molecules and the
    conservative rate over all 1,000 requests.}
    \label{tab:geom_posebusters}
    \scriptsize
    \setlength{\tabcolsep}{5pt}
    \begin{tabular}{@{}lrrrr@{}}
        \toprule
        & \multicolumn{2}{c}{Released EDM} & \multicolumn{2}{c}{\mosaic{}} \\
        Check & Valid & All & Valid & All \\
        \midrule
        Connected atoms & 46.90 & 40.9 & 82.14 & 80.0 \\
        Bond lengths & 88.07 & 76.8 & 98.46 & 95.9 \\
        Bond angles & 91.97 & 80.2 & 98.46 & 95.9 \\
        Ring planarity & 99.08 & 86.4 & 99.79 & 97.2 \\
        Double-bond planarity & 55.28 & 48.2 & 72.07 & 70.2 \\
        Internal energy & 94.52 & 81.1 & 98.46 & 95.9 \\
        Steric clashes & 84.75 & 73.9 & 95.28 & 92.8 \\
        All seven checks & 19.00 & 16.3 & 54.11 & 52.7 \\
        \bottomrule
    \end{tabular}
    \vspace{2pt}

    \begin{minipage}{0.96\linewidth}
    \footnotesize Rates are percentages. Standard check denominators are 872
    strict-valid EDM molecules and 974 strict-valid \mosaic{} molecules.
    \end{minipage}
\end{table}

\paragraph{Matched support-and-prior ablation.}
Table~\ref{tab:geom_support_ablation} gives the complete results for the
controlled Stage-2 comparison summarized in
Table~\ref{tab:molecular_support_ablation_main}. Both models are trained from
scratch for 300,000 updates with the same architecture, dataset, optimizer,
schedule, seed, and fixed CountDiT Stage 1. They are evaluated on the same
1,000 requested atom counts and random seeds. The global model collapses the
predicted slots to one center and uses a Gaussian coordinate prior; the local
model retains the predicted cell anchors and uses a uniform cell-wise prior.

\begin{table}[H]
    \centering
    \caption{Matched support-and-prior ablation on GEOM-Drugs. Brackets show
    bootstrap 95\% intervals over requests; PB-valid is conditional on strict
    validity.}
    \label{tab:geom_support_ablation}
    \small
    \setlength{\tabcolsep}{4pt}
    \resizebox{\linewidth}{!}{%
    \begin{tabular}{@{}lrr@{}}
        \toprule
        Metric & Global center & Local support \\
        \midrule
        Atom stability$^{L}$ & 74.13 [73.70, 74.58] & \textbf{83.44} [82.99, 83.91] \\
        Full-graph connectivity$^{F}$ & 25.6 [23.0, 28.5] & \textbf{67.5} [64.6, 70.3] \\
        Full-graph connected-valid$^{F}$ & 24.3 [21.7, 27.1] & \textbf{65.9} [63.0, 68.8] \\
        Strict RDKit validity$^{R}$ & 63.2 [60.4, 66.1] & \textbf{97.3} [96.2, 98.2] \\
        Strict connectivity$^{R}$ & 33.4 [30.4, 36.4] & \textbf{70.5} [67.7, 73.2] \\
        Charge-marginalized stability$^{C}$ & 52.3 [49.3, 55.4] & \textbf{96.0} [94.7, 97.1] \\
        PB-valid$^{P}$ & 11.37 & \textbf{42.70} \\
        Seconds per molecule & \textbf{0.475} & 0.507 \\
        Connected-valid molecules per second & 0.512 & \textbf{1.299} \\
        \bottomrule
    \end{tabular}
    }
\end{table}

\subsection{Spatial group steering diagnostics}
\label{app:molecular_steering}

Both stages of the molecular cascade use classifier-free guidance. The
end-to-end study fixes the trained Stage-2 model at $\gamma_2=1$ and varies
Stage-1 guidance on eight held-out multi-group prompts containing 30 requested
group instances. Table~\ref{tab:molecular_steering} shows that stronger Stage-1
guidance improves support recall and requested-group identity while preserving
the requested count and coordinate-screen acceptance.

\begin{table}[H]
    \centering
    \caption{End-to-end steering with Stage-2 guidance fixed at
    $\gamma_2=1$. Acceptance is $7/8$ and count MAE is zero at every setting.}
    \label{tab:molecular_steering}
    \small
    \setlength{\tabcolsep}{9pt}
    \begin{tabular}{@{}crrr@{}}
        \toprule
        Stage-1 guidance $\gamma_1$ & Support recall & Type $\mathrm{F}_1$ & Spatial recall \\
        \midrule
        0 & 0.967 & 0.491 & 0.400 \\
        1 & 1.000 & 0.588 & 0.433 \\
        2 & 1.000 & 0.667 & 0.433 \\
        \bottomrule
    \end{tabular}
\end{table}

To isolate the Stage-2 response, we also evaluate guidance on eight
fixed reference supports. Increasing $\gamma_2$ from zero to one raises type
recall from $0.433$ to $0.733$, type precision from $0.520$ to $0.917$, and
spatial recall from $0.333$ to $0.633$, while reducing prompt-center MAE from
$3.359$ to $1.855\,\text{\AA}$.

\section{Polycrystal generation metrics}
\label{app:polycrystal_metrics}

Table~\ref{tab:local_morphology_metrics} reports the complete metrics for the
controlled $4^3$ bicrystal, tricrystal, and eight-grain maps shown in
Figure~\ref{fig:polycrystal_control}(c).

\begin{table}[H]
    \centering
    \caption{Controlled local morphology generation. RDF W1 is measured
    against the source polycrystal.}
    \label{tab:local_morphology_metrics}
    \small
    \setlength{\tabcolsep}{4pt}
    \begin{tabular}{@{}lrrrrr@{}}
        \toprule
        Geometry & States & Agreement & Adj. P/R & FCC & RDF W1 (\AA{}) \\
        \midrule
        Bicrystal & $2/2$ & 0.958 & $1.000/1.000$ & 0.293 & 0.0398 \\
        Tricrystal & $3/3$ & 0.953 & $1.000/1.000$ & 0.271 & 0.0390 \\
        Eight-grain & $8/8$ & 0.862 & $0.955/0.955$ & 0.203 & 0.0442 \\
        \bottomrule
    \end{tabular}
\end{table}

The three large morphology maps contain 41, 42, and 44 nominal orientation
labels, of which 26, 30, and 30 are recovered, respectively.
Table~\ref{tab:million_atom} reports requested and recovered orientation-state
counts separately; these are not counts of connected grains. The generated structures
preserve the principal radial shell positions of the source polycrystal, with
global RDF distances of $0.059$--$0.062$\,\AA{}.

\begin{table}[H]
    \centering
    \caption{Million-atom nickel generation. RDF W1 is measured against the
    source polycrystal.}
    \label{tab:million_atom}
    \small
    \setlength{\tabcolsep}{3.8pt}
    \begin{tabular}{@{}lrrrrrrr@{}}
        \toprule
        Map & Requested & Recovered & Atoms & Agreement & Adj. P/R & FCC & RDF W1 \\
        \midrule
        A & 41 & 26 & $1{,}103{,}842$ & $0.817$ & $0.938/0.409$ & $0.232$ & $0.0592$ \\
        B & 42 & 30 & $1{,}116{,}011$ & $0.874$ & $0.972/0.541$ & $0.263$ & $0.0592$ \\
        C & 44 & 30 & $1{,}087{,}696$ & $0.717$ & $0.867/0.454$ & $0.191$ & $0.0621$ \\
        \bottomrule
    \end{tabular}
\end{table}

\section{Additional condensed-matter benchmarks}
\label{app:broader_condensed}

\subsection{Chemical disorder}

We evaluate \mosaic{} on the Fe$_{60}$Ni$_{20}$Cr$_{20}$ tasks from DISMAI
\citep{yong2024dismai}. These structures use fixed FCC sites and composition,
isolating the generation of short-range chemical order. Each condition contains
1,000 generated structures and is evaluated with the published
cluster-correlation distance $d_{\mathrm{cluster}}$. Relative to the benchmark's
random-alloy reference, \mosaic{} reduces this distance by $35.9$--$69.6\%$
(Table~\ref{tab:dismai_alloy}).

\begin{table}[H]
    \centering
    \caption{Fe$_{60}$Ni$_{20}$Cr$_{20}$ chemical-ordering results on DISMAI.
    Lower $d_{\mathrm{cluster}}$ is better. Intervals are 95\% sampling
    intervals over 1,000 generated structures.}
    \label{tab:dismai_alloy}
    \small
    \setlength{\tabcolsep}{3.8pt}
    \begin{tabular}{@{}lrrrrr@{}}
        \toprule
        Condition & \mosaic{} & 95\% interval & Random reference & Reduction & Unique \\
        \midrule
        300 K, narrow & 0.0234 & [0.0223, 0.0245] & 0.0604 & 61.3\% & 1000/1000 \\
        300 K, wide   & 0.0200 & [0.0189, 0.0211] & 0.0658 & 69.6\% & 1000/1000 \\
        1500 K, narrow& 0.0198 & [0.0189, 0.0206] & 0.0308 & 35.9\% & 1000/1000 \\
        1500 K, wide  & 0.0202 & [0.0193, 0.0210] & 0.0338 & 40.3\% & 1000/1000 \\
        \bottomrule
    \end{tabular}
\end{table}

\subsection{Off-lattice amorphous structures}

The DISMAI benchmark also includes amorphous silicon, where no crystalline
support map is available. All 64 generated structures complete the official
SOAP--GAP relaxation. Table~\ref{tab:dismai_asi} reports the resulting motif,
radial-distribution, and bond-angle distances alongside the published CDVAE and
DiffCSP values.

\begin{table}[H]
    \centering
    \caption{Amorphous-silicon results on DISMAI. Lower distances are better;
    failure denotes structures that did not complete the official relaxation
    and evaluation pipeline.}
    \label{tab:dismai_asi}
    \small
    \setlength{\tabcolsep}{5pt}
    \begin{tabular}{@{}lrrrr@{}}
        \toprule
        Model & Motif distance & RDF distance & Angle distance & Failure \\
        \midrule
        CDVAE & 0.0402 & 0.392 & 0.00332 & 0\% \\
        DiffCSP & 0.0647 & 1.390 & 0.01030 & 0\% \\
        \mosaic{} (64 structures) & 0.1179 & 1.756 & 0.01574 & 0\% \\
        \bottomrule
    \end{tabular}
\end{table}

\subsection{Heterogeneous interfaces}

We additionally evaluate the five-species
Li$_3$ScCl$_6$(100)--LiCoO$_2$(110) interface from DISMAI. The complete
structure is generated without copying slab atoms or prescribing atomic
coordinates. A support restriction derived from the training structures removes
empty cell regions while retaining the full interface composition. Of 1,000
generated structures, 621 complete the official M3GNet relaxation and
coordination analysis. Table~\ref{tab:dismai_interface} reports the resulting
coordination-fingerprint distances.

\begin{table}[H]
    \centering
    \caption{Solid--solid interface results on DISMAI after the official
    M3GNet relaxation. Entries are Euclidean coordination-fingerprint
    distances; lower is better.}
    \label{tab:dismai_interface}
    \scriptsize
    \setlength{\tabcolsep}{3.2pt}
    \begin{tabular}{@{}lrrrrr@{}}
        \toprule
        Model & Li & Co & Sc & All cations & Failure \\
        \midrule
        CDVAE & 0.0664 & 0.0427 & 0.2010 & 0.0547 & 0.0\% \\
        DiffCSP & 0.0439 & 0.0272 & 0.0954 & 0.0370 & 7.2\% \\
        CrysTens & 0.0202 & 0.0141 & 0.0972 & 0.0213 & 55.3\% \\
        CryinGAN & 0.0538 & 0.0274 & 0.0888 & 0.0426 & 9.0\% \\
        \mosaic{} & 0.3951 & 0.4571 & 0.5801 & 0.3529 & 37.9\% \\
        \bottomrule
    \end{tabular}
\end{table}

\section{Additional local-structure diagnostics}
\label{app:local_structure}

\begin{figure}[H]
    \centering
    \includegraphics[width=\linewidth]{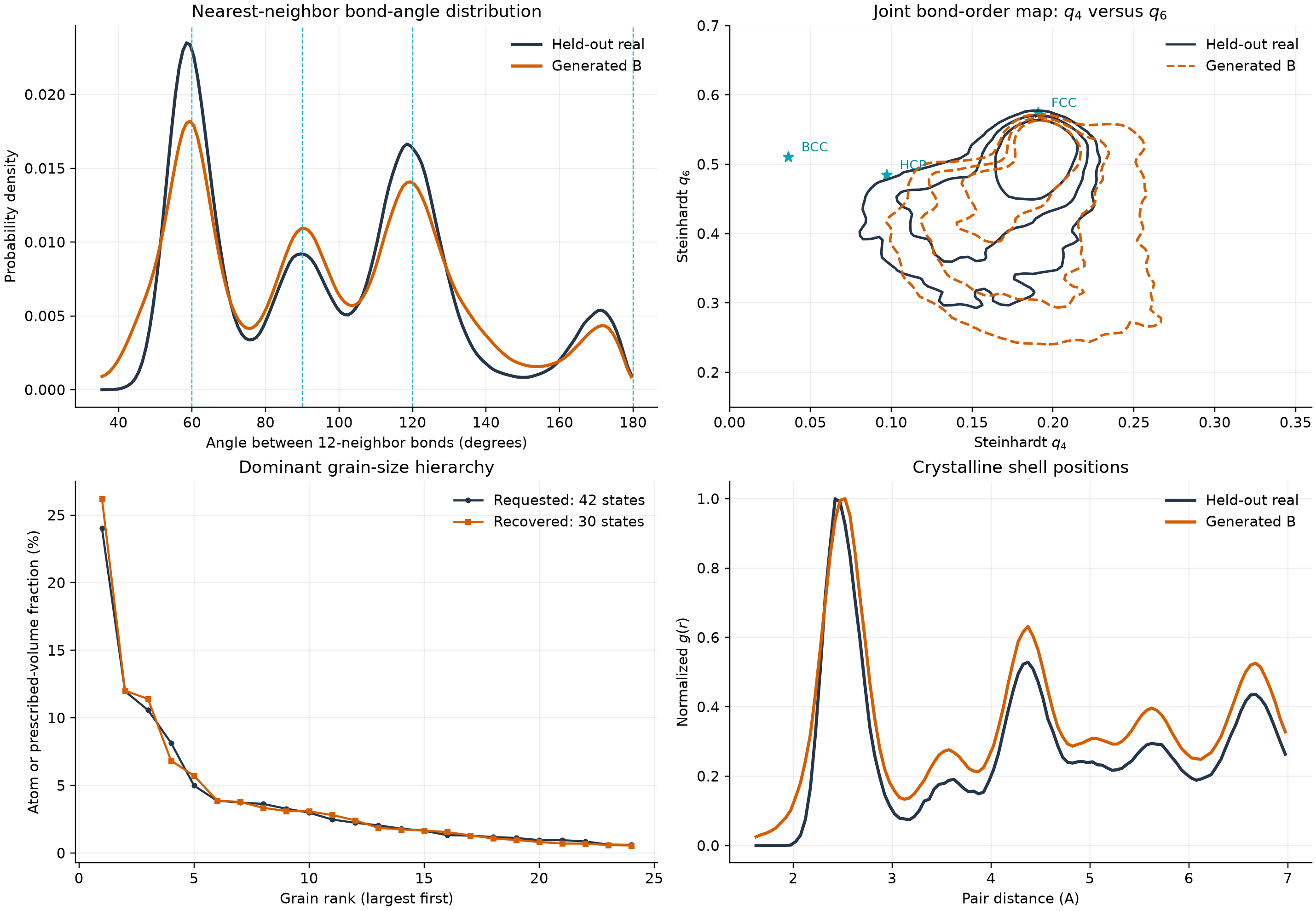}
    \caption{Additional local-structure diagnostics for the representative
    42-state morphology. Generated Map B is compared with held-out source
    material using nearest-neighbor bond angles, the joint Steinhardt
    $(q_4,q_6)$ distribution, dominant grain-size fractions, and radial shell
    positions.}
    \label{fig:local_order_diagnostics}
\end{figure}

Figure~\ref{fig:local_order_diagnostics} provides distributional diagnostics for
the representative 42-state morphology (Map B). The upper panels compare the
nearest-neighbor bond-angle distribution and the joint $(q_4,q_6)$ distribution
with held-out source material. The lower panels show the dominant grain-size
hierarchy and the radial shell positions. The generated structure closely
reproduces the principal crystalline shell locations and dominant grain-size
hierarchy, showing accurate local spacing and morphology across scales. The
bond-angle and bond-order distributions provide complementary measures of
local interfacial structure alongside the global RDF and FCC statistics in
Table~\ref{tab:million_atom}.

\begin{figure}[H]
    \centering
    \includegraphics[width=\linewidth]{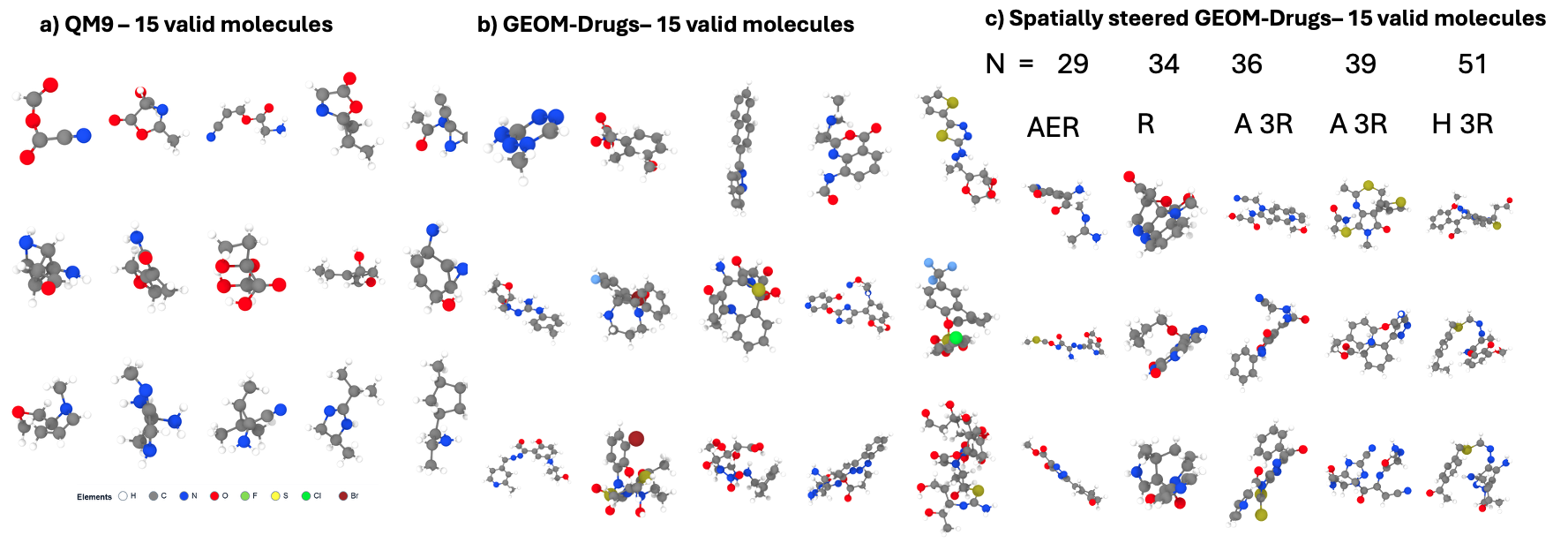}
    \caption{Generated molecular samples. Panels (a) and (b) show 15 valid
    generations from QM9 and GEOM-Drugs, respectively. Panel (c) shows five
    spatially steered GEOM-Drugs prompts at three Stage-1 guidance scales,
    ordered from 0 to 2 from top to bottom. Prompt labels use A for amide,
    E for ester, R for aromatic ring, and H for hydroxyl; a preceding numeral
    gives the requested multiplicity, as in 3R for three aromatic rings.}
    \label{fig:molecular_sample_gallery}
\end{figure}

\begin{figure}[H]
    \centering
    \includegraphics[width=\linewidth]{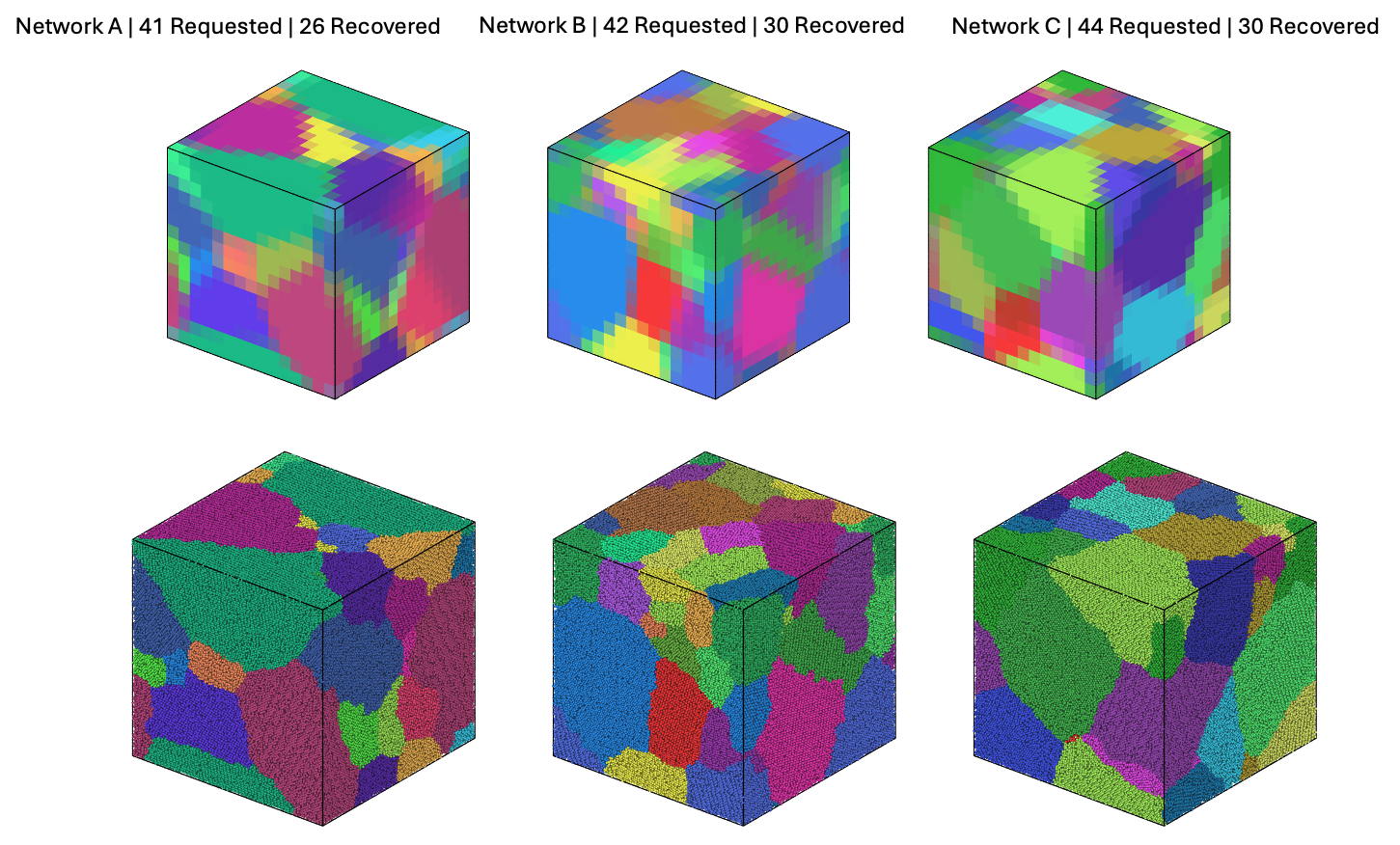}
    \caption{Morphology-conditioned million-atom nickel generation.
    Top: prescribed $15^3$-cell morphology maps for networks A, B, and C.
    Bottom: the corresponding autoregressively generated polycrystals, each
    spanning 24\,nm and containing $1{,}103{,}842$, $1{,}116{,}011$, and
    $1{,}087{,}696$ atoms, respectively. Colors distinguish orientation states.
    The maps specify coarse spatial organization rather than atomistic target
    coordinates. In the headings, ``Requested'' denotes the total number of
    nominal orientation labels in each map (41, 42, and 44), whereas
    ``Recovered'' denotes the number of those states identified in the
    corresponding generated structure (26, 30, and 30). These are
    orientation-state counts, not counts of connected grains. Quantitative results are reported in
    Table~\ref{tab:million_atom}.}
    \label{fig:ni_million_atom_ensemble}
\end{figure}

\end{document}